\documentclass{LMCS}

\def\doi{8(4:7)2012}
\lmcsheading%
{\doi}
{1--39}
{}
{}
{Aug.~\phantom02, 2011}
{Oct.~12, 2012}
{}

\usepackage{pstricks,pst-tree}
\usepackage{ifthen}
\usepackage{enumerate,hyperref}
\usepackage{url}
\usepackage{stmaryrd}

\newcommand {\defin}[1]{\emph{#1}}

\newenvironment{deflist}[1]%
{\begin{list}{}%
{\settowidth{\labelwidth}{#1}%
\setlength{\leftmargin}{\labelwidth}%
\addtolength{\leftmargin}{\labelsep}%
}}%
{\end{list}}

\newenvironment{deflisteng}[1]%
{\begin{list}{}%
{\settowidth{\labelwidth}{#1}%
\setlength{\leftmargin}{\labelwidth}%
\addtolength{\leftmargin}{\labelsep}%
\setlength{\topsep}{0pt}%
\setlength{\itemsep}{0pt}%
}}%
{\end{list}}%

\newcommand {\DL}{\begin{deflist}}
\newcommand {\EDL}{\end{deflist}}
\newcommand {\DLE}{\begin{deflisteng}}
\newcommand {\EDLE}{\end{deflisteng}}

\newcommand {\db}{\displaybreak[0]}

	\newcommand{\ie}{i.e.\ }
	\newcommand{\eg}{e.g.\ }
	\newcommand{\wrt}{w.r.t.\ }
	
	\newcommand{\resp}{resp.\ }
	\newcommand{\esp}{esp.\ }
	
        \newcommand{\famm}{fully abstract model}
        
\newcommand{\no}{\noindent}

\newcommand{\Berry}{\cite{Berry}}
\newcommand{\Laird}{\cite{Laird:sequ}}

\newcommand {\typ}{\colon} 
\newcommand {\fun}{\mathbin{\to}}    
\newcommand {\pro}{\times}    
\newcommand {\lpro}{\langle}    
\newcommand {\rpro}{\rangle}    
\newcommand {\proi}{\pi_1}    
\newcommand {\prot}{\pi_2}    
\newcommand {\mto}{\mathord{\mapsto}}   

\newcommand {\la}{\lambda}
\newcommand {\om}{\omega}
\newcommand {\si}{\sigma}
\newcommand {\ta}{\tau}
\newcommand {\io}{\iota}
\newcommand {\ro}{\rho}
\newcommand {\fp}{\psi}    
\newcommand {\fpsi}{\fp^\si_i}

\newcommand {\FP}{\Psi}    
\newcommand {\FPsi}{\FP^\si_i}
\newcommand {\FPsj}{\FP^\si_j}
\newcommand {\FPti}{\FP^\ta_i}
\newcommand {\FPixn}{(\FP^{\si_1}_i x_1)\ldots(\FP^{\si_n}_i x_n)}
\newcommand {\FPjxn}{(\FP^{\si_1}_j x_1)\ldots(\FP^{\si_n}_j x_n)}
\newcommand {\sinio}{\si_1 \fun \ldots \fun \si_n \fun \io}
\newcommand {\transi}{\typ \si(i,j)}

\newcommand {\vecx}{\vec{x}}
\newcommand {\vecy}{\vec{y}}

\newcommand {\program}[1]{\mbox{\textup{\textsf{#1}}}}

\newcommand {\0}{\program{0}}        
\newcommand {\1}{\program{1}}        
\newcommand {\2}{\program{2}}        
\newcommand {\bo}{\bot}       
\newcommand {\Om}{\bot}       
\newlength{\botw}
\newcommand{\bz}{\makebox[\botw]{\0}}  
\newcommand{\bi}{\makebox[\botw]{\1}}  
\newcommand{\bt}{\makebox[\botw]{\2}}  
\newcommand{\ph}[1]{\phantom{#1}}
\newcommand{\tsp}{\;\;}                

\newcommand {\pif}{\mathop{\program{if}}}
\newcommand {\pzero}{\mathop{\program{zero}}}
\newcommand {\ppre}{\operatorname{\program{pre}}}
\newcommand {\psuc}{\mathop{\program{suc}}}
\newcommand {\pthen}{\mathbin{\program{then}}}
\newcommand {\pelse}{\mathbin{\program{else}}}
\newcommand {\case}{\mathop{\program{case}}}
\newcommand {\casei}{\mathop{{\program{case}}_i}}
\newcommand {\casej}{\mathop{{\program{case}}_j}}
\newcommand {\casez}{\mathop{{\program{case}}_0}}
\newcommand {\caseum}{\mathop{{\program{case}}_1}}
\newcommand {\caset}{\mathop{{\program{case}}_2}}
\newcommand {\caseinf}{\mathop{{\program{case}}_\infty}}
\newcommand {\strict}{\mathop{\program{strict?}}}
\newcommand {\str}{\mathop{\program{str}}}

\newcommand {\Y}{\mathord{\program{Y}}}
\newcommand {\PCF}{\mbox{\textnormal{PCF}}}
\newcommand {\ENV}{\mbox{\textnormal{ENV}}} 
\renewcommand {\inf}{\operatorname{inf}}
\newcommand {\apply}{\operatorname{apply}}

\newcommand {\nf}{\operatorname{nf}}

\renewcommand {\approx}{\operatorname{approx}}
\newcommand {\Inj}{\operatorname{Inj}}
\newcommand {\Proj}{\operatorname{Proj}}
\newcommand {\strictify}{\operatorname{strictify}}
\newcommand {\unstr}{\operatorname{unstr}}
\newcommand {\eval}{\operatorname{eval}}
\newcommand {\new}{\operatorname{new}}
\newcommand {\choice}{\operatorname{choice}}
\newcommand {\gt}{\mathop{{\mbox{\textnormal{gt}}}}}
\newcommand {\gti}{\mathop{{\mbox{\textnormal{gt}}}^\si_i}}
\newcommand {\gts}{\mathop{{\mbox{\textnormal{gt}}}^\si}}

\newcommand {\gtj}{\mathop{{\mbox{\textnormal{gt}}}^\si_j}}
\newcommand {\gtz}{\mathop{{\mbox{\textnormal{gt}}}^\si_0}}
\newcommand {\projsi}{\mathop{{\mbox{\textnormal{proj}}}^\si_i}}
\newcommand {\projsj}{\mathop{{\mbox{\textnormal{proj}}}^\si_j}}
\newcommand {\laxn}{\la x_1 \ldots x_n.}

\newcommand {\Mm}{M_1 \ldots M_m}
\newcommand {\An}{A_1 \ldots A_n}
\newcommand {\Bn}{B_1 \ldots B_n}
\newcommand {\Ni}{N_0 \ldots N_i}

\newcommand {\Ns}{N^{\ast}}
\newcommand {\As}{A^{\ast}}
\newcommand {\Bs}{B^{\ast}}

\newcommand {\outrep}[1]{\lceil #1 \rceil}

\newcommand {\lop}{\sqsubseteq_{op}}  
\newcommand {\lsy}{\prec}     
\newcommand {\lsys}{\prec^s}     
\newcommand {\lst}{\leq}      
\newcommand {\glbst}{\wedge}      
\newcommand {\lubst}{\vee}      
\newcommand {\lex}{\sqsubseteq} 
\newcommand {\gex}{\sqsupseteq} 
\newcommand {\glbex}{\sqcap} 
\newcommand {\Glbex}{\bigsqcap} 
\newcommand {\lubex}{\sqcup} 
\newcommand {\Lubex}{\bigsqcup} 
\newcommand {\Lubst}{\bigvee} 
\newcommand {\eq}{\cong}        
\newcommand {\cost}{\uparrow_{\lst}} 
\newcommand {\coh}{\uparrow_{h}} 
\newcommand {\retract}{\unlhd}

\newcommand {\re}{\to}        
\newcommand {\ret}{\to^{\ast}} 
\newcommand {\reby}{\to_{\beta Y}}
\newcommand {\rebyt}{\reby^{\ast}}

\newcommand {\redi}{\downarrow_i}
\newcommand {\redj}{\downarrow_j}

\newcommand {\diver}{\mathord{\uparrow}}   
\newcommand {\conver}{\mathord{\downarrow}}   

\newcommand {\sem}[1]{[\![ #1 ]\!]}  
\newcommand {\tsem}[1]{\mathcal{T}[\![ #1 ]\!]}  
\newcommand {\trace}{\mathcal{T}}
\newcommand {\eclass}[1]{[ #1 ]_{op}}         
\newcommand {\fin}{\mathcal{F}}
\newcommand {\fins}{\fin^\si}
\newcommand {\finsi}{\fin^\si_i}
\newcommand {\finsz}{\fin^\si_0}
\newcommand {\Ds}{D^\si}
\newcommand {\Dt}{D^\ta}
\newcommand {\Dst}{D^{\si\fun\ta}}
\newcommand {\Es}{E^\si}
\newcommand {\Et}{E^\ta}
\newcommand {\Est}{E^{\si\fun\ta}}
\newcommand {\Dsn}{D^{\si_1\fun\dots\fun\si_n\fun\io}}
\newcommand {\down}{\mathord{\downarrow}}
\newcommand {\flub}{\to}
\newcommand {\id}{\mbox{\textnormal{id}}}
\newcommand {\comp}{\circ}

\newcommand {\set}[2]{\{\, #1 \mid #2 \,\}}  
\newcommand {\ifthen}{\Longrightarrow}
\newcommand {\thenif}{\Longleftarrow}
\newcommand {\ifff}{\Longleftrightarrow}
\newcommand {\infer}[2]{$\dfrac{#1}{#2}$}
\newcommand {\sub}{\subseteq}
\renewcommand {\Join}{\bigcup}
\newcommand {\all}[1]{\forall #1 . \;}
\newcommand {\exi}[1]{\exists #1 . \;}

\newlength{\be}
\newlength{\len}
\newlength{\me}
\newlength{\disp}
\newlength{\zeroh}
\newlength{\brackh}
\newlength{\brackd}
\newcommand{\leftb}[3]{%
\setlength{\len}{#2\be}%
\addtolength{\len}{-#1\be}%
\addtolength{\len}{\brackh}%
\addtolength{\len}{\brackd}%
\setlength{\me}{0.5\len}%
\addtolength{\me}{#1\be}%
\addtolength{\me}{-\be}%
\addtolength{\me}{0.35\zeroh}%
\addtolength{\me}{-\brackh}%
\setlength{\disp}{-0.5\len}%
\addtolength{\disp}{0.35\zeroh}%
\raisebox{-\me}{$ #3 \left\{ \rule[\disp]{0pt}{\len} \right. $}}

\newcommand{\rightb}[3]{%
\setlength{\len}{#2\be}%
\addtolength{\len}{-#1\be}%
\addtolength{\len}{\brackh}%
\addtolength{\len}{\brackd}%
\setlength{\me}{0.5\len}%
\addtolength{\me}{#1\be}%
\addtolength{\me}{-\be}%
\addtolength{\me}{0.35\zeroh}%
\addtolength{\me}{-\brackh}%
\setlength{\disp}{-0.5\len}%
\addtolength{\disp}{0.35\zeroh}%
\raisebox{-\me}{$ \left. \rule[\disp]{0pt}{\len} \right\} #3 $}}

\psset{linewidth=0.5pt}
\psset{nodesep=3pt}
\psset{treesep=4mm}
\psset{levelsep=10mm}
\newcommand{\levelsepp}{10mm}

\newboolean{btree}  
\setboolean{btree}{true}
\newboolean{outer}  

\newlength{\varwidth}
\newcommand{\labox}[2]{\settowidth{\varwidth}{$#2$}
                       \makebox[\varwidth][r]{$#1 \, #2$}}

\newcommand{\tz}{\ifthenelse{\boolean{btree}}{\TR{\0}}{\bz}} 
\newcommand{\ti}{\ifthenelse{\boolean{btree}}{\TR{\1}}{\bi}} 
\newcommand{\ttwo}{\ifthenelse{\boolean{btree}}{\TR{\2}}{\bt}} 
\newcommand{\tbo}{\ifthenelse{\boolean{btree}}{\TR{$\ph{\bo}$}}{\Om}} 
\newcommand{\tc}[1]{\ifthenelse{\boolean{btree}}{\TR{$#1$}}{#1}} 

\newcommand{\tx}[4][\levelsepp]
{\ifthenelse{\boolean{btree}}%
            {\pstree[treemode=R,thislevelsep=#1]{\TR{$#2$}}{#3 #4}}%
            {\ifthenelse{\boolean{outer}}%
                        {\setboolean{outer}{false} $\caseum #2  #3  #4$}%
                        {( \caseum #2  #3  #4 )}}%
}

\newcommand{\tlx}[5][\levelsepp]
{\ifthenelse{\boolean{btree}}%
            {\pstree[treemode=R,thislevelsep=#1]{\TR{\labox{#2}{#3}}}{#4 #5}}%
            {\ifthenelse{\boolean{outer}}%
                        {\setboolean{outer}{false} $#2\,\caseum #3  #4 #5$}%
                        {(#2\, \caseum #3  #4  #5 )}}%
}

\newcommand{\tf}[5][\levelsepp]
{\ifthenelse{\boolean{btree}}%
            {\pstree[treemode=R,thislevelsep=#1]%
            {\pstree[treemode=D]{\TR{$#2$}}{#3}}{#4 #5}}%
            {\ifthenelse{\boolean{outer}}%
                        {\setboolean{outer}{false} $\caseum (#2 #3)  #4  #5$}%
                        {( \caseum (#2 #3)  #4  #5 )}}%
}

\newcommand{\tlf}[6][\levelsepp]
{\ifthenelse{\boolean{btree}}%
            {\pstree[treemode=R,thislevelsep=#1]%
            {\pstree[treemode=D]{\TR{\labox{#2}{#3}}}{#4}}{#5 #6}}%
            {\ifthenelse{\boolean{outer}}%
                        {\setboolean{outer}{false}$#2\,\caseum (#3 #4)  #5  #6$}%
                        {(#2\, \caseum (#3 #4)  #5  #6 )}}%
}

\newcommand{\tttlf}[5][\levelsepp]
{\ifthenelse{\boolean{btree}}%
            {\pstree[treemode=R,thislevelsep=#1]%
            {\pstree[treemode=D]{\TR{\labox{#2}{#3}}}{#4}}{#5}}%
            {\ifthenelse{\boolean{outer}}%
                        {\setboolean{outer}{false}$#2\, \casei (#3 #4)  #5 $}%
                        {(#2\, \casei (#3 #4)  #5 )}}%
}

\newlength{\variwidth}
\newlength{\vawidth}
\newlength{\lawidth}
\newcommand{\emptyedge}[2]{}

\newcommand{\ttlf}[5][\levelsepp]
{\settowidth{\variwidth}{$\la #2.\, #3$}%
\settowidth{\vawidth}{$#3$}%
\addtolength{\variwidth}{-0.5\lawidth}%
\addtolength{\variwidth}{-0.5\vawidth}%
\pstree[treemode=R,thislevelsep=\variwidth]{\TR{$\la$}}%
{\pstree[treemode=R,thislevelsep=#1]%
{\pstree[treemode=D]{\TR[edge=\emptyedge]{\labox{#2.}{#3}}}{#4}}%
{#5}}}

\newcommand{\ttlx}[4][\levelsepp]
{\settowidth{\variwidth}{$\la #2.\, #3$}%
\settowidth{\vawidth}{$#3$}%
\addtolength{\variwidth}{-0.5\lawidth}%
\addtolength{\variwidth}{-0.5\vawidth}%
\pstree[treemode=R,thislevelsep=\variwidth]{\TR{$\la$}}%
{\pstree[treemode=R,thislevelsep=#1]%
{\TR[edge=\emptyedge]{\labox{#2.}{#3}}}%
{#4}}}

\newcommand{\longarm}[1]
{\ifthenelse{\boolean{btree}}{\skiplevel{#1}}{#1}}

\newcommand{\treesp}[1]
{\ifthenelse{\boolean{btree}}{\tspace{#1}}{}}

\newcommand{\tree}[1]
{\ifthenelse{\boolean{btree}}{#1}%
            {\typeout{Warning: tree inside string}%
             \setboolean{btree}{true} #1 \setboolean{btree}{false}}%
}

\newcommand{\lstring}[1]
{\ifthenelse{\boolean{btree}}%
            {\setboolean{btree}{false}\setboolean{outer}{true}%
             #1 \setboolean{btree}{true}}%
            {#1}%
} 

\begin{document}
\title[On Berry's Conjectures about the Stable Order in PCF]{On Berry's Conjectures about the Stable Order in PCF}
\author[F.~M\"uller]{Fritz M\"uller}
\address{Saarland University, Department of Computer Science, Campus E1.3, 66123 Saarbr\"ucken, Germany,
 \url{http://rw4.cs.uni-saarland.de/~mueller}}
\email{\texttt{($\la$x.muellerxcs.uni-saarland.de)@} }

\begin{abstract}
PCF is a sequential simply typed lambda calculus language.
There is a unique order-extensional fully abstract cpo-model of PCF, built up from
equivalence classes of terms.
In 1979, G\'erard Berry defined the stable order in this model and proved
that the extensional and the stable order together form a bicpo.
He made the following two conjectures:\\
1) ``Extensional and stable order form not only a bicpo, but a bidomain.''\\
We refute this conjecture by showing that the stable order is not bounded complete,
already for finitary PCF of second-order types.\\
2) ``The stable order of the model 
has the syntactic order as its image:
If $a$ is less than $b$ in the stable order of the model, for finite $a$ and $b$,
then there are normal form terms $A$ and $B$ with the semantics $a$, resp.\ $b$,
such that $A$ is less than $B$ in the syntactic order.''\\
We give counter-examples to this conjecture, again in finitary PCF of second-order
types, and also refute an improved conjecture: There seems to be no 
simple syntactic characterization of the stable order. But we show that Berry's
conjecture is true for unary PCF.

For the preliminaries, we explain the basic fully abstract semantics of PCF in the general setting of
(not-necessarily complete) partial order models (f-models).
And we restrict the syntax to ``game terms'', with a graphical representation.
\end{abstract}

\maketitle
\renewcommand{\sfdefault}{cmss}

\section{Introduction}

PCF is a simple functional programming language, a call-by-name typed lambda calculus
with integers and booleans as ground types, some simple sequential operations on the 
ground types, and a fixpoint combinator.
The concept of PCF was formed by Dana Scott in 1969, see the historical document
\cite{Scott}.
It is used as a prototypical programming language to explore the relationship
between operational and denotational semantics, see the seminal paper of
Gordon Plotkin \cite{Plotkin}.

The (operational) \defin{observational preorder} $M \lop N$
of two terms (of equal type) is defined as:
For all contexts $C[\;]$ of integer type, if $C[M]$ reduces to the integer $n$,
then $C[N]$ also reduces to the same $n$.
The denotational semantics (the model) assigns to every term $M$ an element $\sem{M}$
of a partial order $(D,\lex)$ (usually a complete partial order, cpo) as meaning.
The model is said to be \defin{(order) fully abstract} if the two orders coincide:
$M \lop N \ifff \sem{M} \lex \sem{N}$.
The standard model of Scott domains and continuous functions is adequate (\ie the
direction $\thenif$ of the coincidence), but not fully abstract,
because the semantic domains contain finite elements that are not expressible as terms,
like the parallel or function.
First Robin Milner \cite{Milner} constructed in 1977 a unique fully abstract order-extensional cpo-model 
of PCF that can be built up from equivalence classes of terms by some ideal
completion. The problem to construct a fully abstract model of PCF that does not use the syntax of terms
(the ``full abstraction problem'') was the driving force of the subsequent developments,
see also the handbook article \cite{Ong}.

In 1979 G\'erard Berry published his PhD thesis \cite{Berry} with the
translated title ``Fully abstract and stable models of typed lambda-calculi'',
which is the main basis of our work.
In order to sort out functions like the parallel or from the semantic domains,
to get ``closer'' to the fully abstract model, he gave the definition of stable
function:
A function $f$ is \defin{stable} if for the computation of some finite part of the
output a deterministic minimal part of the input is needed.
In the case that there are only finitely many elements smaller than a finite element,
this definition is equivalent to the definition of a \defin{conditionally 
multiplicative function} $f$: If $a$ and $b$ are compatible, then
$f(a \glbex b) = f a \glbex f b$.
To make the operation of functional application of stable functions itself stable,
Berry had to replace the pointwise order of functions, the extensional order,
by the new stable order: Two functions are in the \defin{stable order}, $f \lst g$,
if for all $x \lst y$: $f x = f y \glbex g x$.
This entails the pointwise order, but it demands in addition that $g$ must not
output some result for input $x$ that $f$ outputs only for greater $y$.

Side remark: Stability is a universal concept that was independently (re)discovered in
many mathematical contexts.
So Jean-Yves Girard found it in the logical theory of dilators
and then transferred it to domain theory (qualitative domain, coherence space)
to give a model of polymorphism (system F) \cite{Girard:F},
thereby independently reinventing Berry's stable functions and stable order,
see also the textbook \cite{Girard/Taylor}, chapter 8 and appendix A.
For a general theory of stability and an extensive bibliography see \cite{Taylor:stable}.

Now Berry had a model (of PCF) of stable functions with the stable order.
But this model did not respect the old (pointwise) extensional order of the
standard model and so had new unwanted elements not contained in the standard model.
To get a proper subset of the standard model, he introduced bicpo models.
A \defin{bicpo} is a set with two orders, an extensional and a stable one, 
both forming cpos and being connected in some way. 
He augmented Milner's fully abstract cpo model by the stable order and proved
that it consists of bicpos and its functions are conditionally multiplicative.
In section \ref{s:semantics} we show in addition that its stable order forms
stable bifinite domains and therefore its functions are also stable and can be
represented by \defin{traces}, \ie sets of tokens (or events) like in \cite{Curien/Plotkin}.
E.g.\ the function $\sem{\la f. \pif (\pzero (f \0)) \pthen \0 \pelse \Om}$
can be represented by the trace consisting of the tokens $\{\0 \mto \0\} \mto \0$
and $\{\bo \mto \0\} \mto \0$.
Functions are in the stable order, $f \lst g$, iff the trace of $f$ is a subset of the
trace of $g$.

\bigskip
In his thesis Berry made the following two conjectures that we refute:

\no \textbf{1)} ``Extensional and stable order in the fully abstract cpo-model of PCF
form not only a bicpo, but a bidomain.''

This would mean (among other things) that the stable order is bounded complete and distributive.
We give counter-examples in finitary PCF 
of second-order types to this conjecture.
The idea is that the stable lub of two stably bounded elements $a$ and $b$ may entail
a new token that was not present in $a$ or $b$.
This new token must be used in the syntax to separate a subterm denoting $a$
from a subterm denoting $b$ that cannot be unified in a common term.
Therefore distributivity is not fulfilled, stable lubs are not taken pointwise.
And worse: There may be a choice between different new tokens to be entailed,
then there is a choice between different minimal stable upper bounds of $a$ and $b$,
but there is no stable lub.
The minimal stable upper bounds are pairwise stably incompatible, and the extensional lub
$a \lubex b$ is one of them.

\bigskip
\no \textbf{2)}
The extensional order of the \famm\ coincides with the (syntactic) observational preorder.
This leads to the question: Is there a syntactic characterization also for the
stable order? Berry made the conjecture:\\
``The stable order of the model 
has the syntactic order as its image:\\
If $a \lst b$ in the stable order, for finite $a$ and $b$,
then there are normal form terms $A$ and $B$ with $\sem{A}=a$ and $\sem{B}=b$,
such that $A \lsy B$ in the syntactic order.''

Berry proved the converse direction: If $A \lsy B$, then $\sem{A} \lst \sem{B}$,
and proved the conjecture for first-order types.

Our simplest counter-example to this conjecture is a situation of four terms
$A \lsy B \eq C \lsy D$, where $\eq$ is observational equivalence, so that 
$\sem{A} \lst \sem{D}$,
but there is no way to find terms $A'\eq A$, $D'\eq D$ with $A' \lsy D'$.
The elimination of some token of $D$ depends on the prior elimination of some other
token, so that two $\lsy$-steps are necessary to get from $D$ down to $A$.

We further give examples where such a chain of $\lsy$-steps (with intermediate 
$\eq$-steps) of any length is necessary.
This proposes an improved conjecture, the ``chain conjecture'':
Instead of $A\lsy B$ we demand the existence of a chain between $A$ and $B$.
But we also refute this conjecture.
Although stable order and syntactic order are connected, there seems to be no
simple syntactic characterization of the stable order in PCF.

\medskip
All our counter-examples for both conjectures are in finitary PCF of second-order types.
They all share a common basic idea:
We have a term $M\typ (\io\fun\io\fun\io)\fun\io$ with two tokens (among others) which are in
the simplest form like the tokens $\{\bo\bo\mto\0\}\mto\0$
and $\{\bo\0\mto\0, \1\1\mto\1\}\mto\0$.
The function call that realizes $\bo\bo\mto\0$ \resp $\bo\0\mto\0$
is at the top level of $M$,
the function call for $\1\1\mto\1$ is nested below.
We want to eliminate the token 
$\{\bo\bo\mto\0\}\mto\0$.
For this the function call for $\1\1\mto\1$ must be ``lifted'' to the top level,
but this is not possible due to other tokens of $M$ that have to stay.

The necessary ingredients for the counter-examples are: at least second-order type
with some functional parameter of arity at least $2$,
at least two different ground values $\0$ and $\1$, and the need for nested function calls.

If we restrict the calculus to a single ground value $\0$, we get unary PCF, and in this case
both of Berry's conjectures are true:
The \famm\ is a bidomain, in fact it is the standard semantical bidomain construction,
proved by Jim Laird in \cite{Laird:sequ}.
And we prove that the syntactic order is the image of the stable order,
using Laird's proof that every type in unary PCF is a definable retract of some 
first-order type.

The need for nested function calls is the result of a ``restriction'' of PCF:
There is no operator to test if a function demands a certain argument,
so that this information could be used in an if-then-else.
Jim Laird has shown that in a language with such control operators (SPCF) nested function calls
can be eliminated, and also every type of SPCF is a definable retract of a first-order type
\cite{Laird:nest}.
Therefore I am convinced, though I do not prove it here, that also for SPCF the syntactic order 
is the image of the stable order.

\medskip
The above mentioned ``restriction'' of PCF is generally the reason for many
irregularities of the semantics of PCF and the difficulty of the full abstraction problem.
An important result is the undecidability of finitary PCF \cite{Loader}.
This means that the observational equivalence of two terms of finitary PCF is undecidable,
and also the question whether there is a term for a functional value table.
As remarked in the introduction to \cite{Curien/Plotkin},
this result restricts the possible fully abstract models of PCF to be not ``finitary''
in some sense.
There have been several solutions for semantical fully abstract models of PCF:
A model of continuous functions restricted by Kripke logical relations \cite{OHearn/Riecke},
and game semantics \cite{Abramsky/Jagadeesan,Hyland/Ong,Nickau}.
In game semantics a term of PCF is modeled by a strategy of a game,
\ie by a process that performs a dialogue of questions and answers with the environment, the opponent.
These strategies are still intensional; the \famm\ is formed by a quotient, the extensional
collapse.
The strategies can be identified with
PCF B\"ohm trees of a certain normal form,
see also \cite[section 6.6]{Amadio/Curien}.
We call these B\"ohm trees ``game terms'' and prove that it is sufficient to
formulate all our results in the realm of game terms,
\esp that if two terms are syntactically ordered, then there are equivalent game terms
so ordered.
This simplifies the proofs of the counter-examples.
We also introduce a graphical notation for game terms that facilitates the handling of larger examples.

It was an open problem whether the game model is isomorphic to Milner's fully abstract cpo-model,
\ie whether its domains are cpos.
This problem was solved by Dag Normann \cite{Normann}: 
Its domains are not cpos, \ie there are directed sets that have no lub.
Then Vladimir Sazonov made a first attempt to build a general theory for these
non-cpo domains \cite{Sazonov:models,Sazonov:natural,Normann/Sazonov}.
His main insight was that functions are continuous only with respect to certain lubs of directed sets
that he calls ``natural lubs''; these are the hereditarily pointwise lubs.

We want to place our results in the context of these new, more general models.
For the semantic preliminaries we
give a simple definition of a set of well-behaved (not-necessarily complete) partial order
fully abstract models of PCF:
These \emph{f-models} are sets of ideals of finite elements, such that application is defined 
and every PCF-term has a denotation.
Sazonov's natural lubs correspond to  our \emph{f-lubs}, which are defined with respect to the finite
elements.

\medskip
I found the counter-example to Berry's second conjecture around the year 1990,
but did not yet publish it.
As far as I know, nobody else tackled Berry's problems.
The reason for this seems to be that they were simply forgotten.
The stable order in the \famm\ was never explored after Berry; a reason may be that
he never prepared a journal version of his thesis, which is not easily accessible.
The recommended introduction to our subject is the report ``Full abstraction for sequential
languages: The state of the art'' \cite{Berry/Curien},
which contains the thesis in condensed form, but lacks most proofs.
There is also an article \cite{Berry:old} published by Berry before his thesis,
which is not recommended, because section 4.5 (bidomains) is wrong
(different definition of bidomain, the first conjecture is stated as theorem).
An excellent general introduction to domains, stability and PCF (and many other things) is the textbook
\cite{Amadio/Curien}. 
But for the stable order in the \famm\ of PCF the only detailed source remains Berry's thesis.

Here is the structure of the paper. The counter-examples are given in the order
of their discovery, \ie in the order of increasing complexity.
\begin{enumerate}[1.]
\setcounter{enumi}{1}
\item Syntax of PCF.
\item Semantics of PCF: non-complete partial order f-models:\\
We introduce f-models as general (not-necessarily complete) partial order fully abstract
models of PCF and give the properties of the stable order in this general context.
(The order-extensional fully abstract cpo-model of PCF is a special case.)
\item Game terms:\\
We describe the construction of game terms by the finite projections
and give a graphical notation for game terms.\\
The expert who is interested only in the counter-examples may skip the introductory sections 2-4;
reading only the definition of game terms and their graphical notation at the beginning of section 4.
\item The syntactic order is not the image of the stable order:\\
We prove Berry's second conjecture for first-order types,
give a counter-example in a second-order type (a chain of length $2$),
and prove the existence of chains of any least length.
\item The stable order is not bounded complete: no bidomain:\\
We prove Berry's first conjecture for first-order types.
In a second-order type we give an example of a stable lub that does not fulfill distributivity,
and an example of two stably bounded elements without stable lub.
\item Refutation and improvement of the chain-conjecture:\\
We refute the improved second conjecture that the stable order entails a chain of terms.
We propose in turn an improvement of the chain conjecture,
based on the complementary syntactic relation of strictification.
\item Unary PCF:\\
We prove Berry's second conjecture for unary PCF,
with the aid of Jim Laird's definable retractions from any type to some first-order type \Laird.
\item Outlook.
\end{enumerate}

\section{Syntax of PCF} \label{s:syntax}

In this section we give the syntactic definitions of PCF \cite{Plotkin,Berry/Curien,Amadio/Curien}.
The programming language PCF is a simply typed lambda calculus with arithmetic
and fixpoint operators.
It usually comes with two ground types $\io$ (integers) and $o$ (booleans).
We simplify the language and use only the ground type $\io$ (integers);
the booleans are superfluous and can be coded as integers,
the intensional structure of the terms stays the same.

\bigskip
\no The \textbf{types} are formed by $\io$ and function types $\si\fun\ta$ for types $\si$ and $\ta$.
\bigskip

\no The typed \textbf{constants} are:\\
$\0, \1, \2, \ldots \typ \io$, the integers;\\
$\psuc, \ppre \typ \io\fun\io$, successor and predecessor function;\\
$\pif \_ \pthen \_ \pelse \_ \typ \io\fun\io\fun\io\fun\io$, this conditional tests if the first argument is $\0$.\\
(We write \eg $\pif x \pthen y$ for the application of this function to only two arguments.)
\bigskip

\no The PCF \textbf{terms} comprise the constants and the typed constructs by the following rules:\\
$\Om^\si \typ \si$ for any type $\si$, the undefined term.\\
$x^\si \typ \si$ for any variable $x^\si$.\\
If $M\typ \ta$, then $\la x^\si.M \typ \si\fun\ta$, lambda abstraction.\\
If $M\typ \si \fun \ta$ and $N\typ\si$, then $MN \typ \ta$, function application.\\
If $M\typ \si\fun \si$, then $\Y M \typ\si$, $\Y$ is the fixpoint operator.
\bigskip

$\PCF^\si$ is the set of all $\PCF$ terms of type $\si$,
and $\PCF^\si_c$ is the set of the closed terms of these.\\
Type annotations of $\Om$ and of variables will often be omitted.\\
We use the (semantic) symbol $\Om$ also as syntactic term, instead of the usual $\Omega$.\\
We define the \defin{syntactic order} $\lsy$ (also called $\Om$-match order in the literature) on terms of the same type:\\  
$M \lsy N$ iff $N$ can be obtained by replacing some occurrences of $\Om$ in $M$ by terms.
\bigskip

\no The \textbf{reduction rules} are (where $n$ is a variable for integer constants):\\
$(\la x.M) N \re M[x:=N]$, the usual $\beta$-reduction;\\
$\Y M \re M(\Y M)$;\\
$\psuc n \re (n+1)$;\\
$\ppre n \re (n-1)$, for $n\geq \1$;\\
$\pif \0 \pthen M \pelse N \re M$;\\
$\pif n \pthen M \pelse N \re N$, for $n\geq \1$.
\bigskip

The \defin{reduction relation} $\re$ is one step of reduction by these rules in any term context.
It is confluent.
$\ret$ is the reflexive, transitive closure of $\re$.

A \defin{program} is a closed term of type $\io$.\\
The \defin{operational (observational) preorder} $\lop$ on terms of the same type is defined as:\\
$M \lop N$ ($M$ is \defin{operationally less defined than} $N$) iff\\
$P[M] \ret n$ implies $P[N] \ret n$ for all contexts $P[\;]$ such that $P[M]$ and $P[N]$
are both programs.\\
The \defin{operational equivalence} is defined as: 
$M \eq N$ iff $M \lop N$ and $N \lop M$.

\section{Semantics of PCF: non-complete partial order f-models} \label{s:semantics}

This section gives an exposition of the fully abstract semantics of PCF with the stable order,
as far as it is needed to understand the results of this paper.
The proofs are omitted, as they are easy and/or already known in some form.

The order-extensional fully abstract cpo-model of PCF was first constructed by Robin Milner
\cite{Milner} based on terms of an SKI-combinator calculus.
Later G\'erard Berry's thesis \cite{Berry} constructed this model based on the proper $\la$-terms.
This model is the ideal completion of the finite elements;
every directed set has a lub.

Then came the fully abstract game models of PCF \cite{Abramsky/Jagadeesan,Hyland/Ong,Nickau}.
The elements of these models can be represented by the (infinite) B\"ohm trees of PCF.
It was an open problem whether the game model is isomorphic to Milner's model,
\ie whether its domains are cpos.

This problem was solved by Dag Normann \cite{Normann}: 
Its domains are not cpos, \ie there are directed sets that have no lub.
Then Vladimir Sazonov made a first attempt to build a general theory for these
non-cpo domains \cite{Sazonov:models,Sazonov:natural,Normann/Sazonov}.
His main insight was that functions are continuous only with respect to certain lubs of directed sets
that he calls ``natural lubs''; these are the hereditarily pointwise lubs.

We want to place our results in the context of these new, more general models.
Therefore we
give a simple definition of a set of well-behaved (not-necessarily complete) partial order
fully abstract models of PCF:
These \emph{f-models} are sets of ideals of finite elements, such that application is defined 
and every PCF-term has a denotation.
Sazonov's natural lubs correspond to  our \emph{f-lubs}, which are defined with respect to the finite
elements.

We state the usual properties for these f-models; the essence of their proofs is already
contained in Berry's construction.
Our aim is the definition of the stable order and of conditionally multiplicative (cm) functions.
All functions in f-models are cm.
We can further show, in addition to Berry, that the domains have property I under the stable order and therefore
the functions are stable and we can work with their traces.

\bigskip

We need the following PCF terms, the \emph{finite projections}
 on type $\si$ of \emph{grade} $i$,\\ $\FPsi :\si\fun\si$:
\begin{align*}
\FP^\io_i  & = \la x^\io. \pif x \pthen \0 \pelse \pif \ppre^1 x \pthen \1 \pelse
                 \ldots \pif \ppre^i x \pthen i \pelse \bo  \\
\FP^{\si\fun\ta}_i  & = \la f^{\si\fun\ta}. \la x^\si. \FPti(f(\FPsi x))
\end{align*}
We also need the following terms for the glb functions on all types,
$\inf^\si \typ \si\fun\si\fun\si$, here in a liberal syntax:
\begin{alignat*}{2}
\inf^\io  &= \la x^\io y^\io. \pif x = y \pthen x \pelse \Om \\
          &= \la x^\io y^\io. \pif x \pthen \pif y \pthen \0 \pelse \Om \\
         &\phantom{= \la x^\io y^\io. \pif x {}} 
                                 { } \pelse \psuc(\inf^\io(\ppre x)(\ppre y)) \\
\inf^{\si\fun\ta} &= \la f^{\si\fun\ta} g^{\si\fun\ta}. \la x^\si. \inf^\ta(f x)(g x)
\end{alignat*}
When applied to a closed term $M\typ\si$, the function term $\FPsi$ serves as a ``filter''
that lets only pass integer values $\leq i$ as input or output to $M$.
This serves to define the finite elements of the intended model.

\begin{defi}
A term $M\typ \si$ is a \defin{finite term of grade} $i$ if it is closed and $M\eq \FPsi M$.\\
$\finsi = \set{\eclass{\FPsi M}}{M \in \PCF^\si_c}$ is the set of \defin{finite elements of grade} $i$
of type $\si$,\\
where $\eclass{X}$ is the equivalence class of term $X$ under the operational equivalence $\eq$.\\
$\fins = \Join_i \finsi$ is the set of \defin{finite elements} of type $\si$.
\end{defi}

The finite elements are partially ordered by the extension of the operational preorder $\lop$
to equivalence classes.\\ 
An \defin{ideal} of finite elements of type $\si$ is a set $S\sub \fins$ such that: $S \neq \emptyset$ and\\
$a,b \in S \ifthen \exi{c \in S} a\lop c \text{ and } b\lop c$,\\
 and
$a\in S, b \in \fins \text{ and } b\lop a \ifthen b\in S$.\\
$I(\fins)$ is the set of ideals of finite elements of type $\si$.\\
There is an operation $\apply$ on ideals of finite elements. 
For $f\in I(\fin^{\si\fun\ta})$, $d\in I(\fins)$:
\[\apply(f,d) = \down\set{f'd'}{f'\in f, d'\in d} \in I(\fin^\ta),\]
where $f'd' = \eclass{M N}$ for $M\in f'$, $N\in d'$.
$\apply(f,d)$ is simply written $f d$.\\
From now on $a\in \fins$ is identified with the ideal $\down\{a\}$, the downward closure \wrt $\lop$ of $\{a\}$.
So we have the embedding $\fins \sub I(\fins)$.

\begin{defi}
An \defin{f-model} of PCF (``f'' means: based on finite elements) is a collection of $D^\si \sub I(\fins)$ for every type $\si$,
each $D^\si$ ordered by inclusion $\sub$ written $\lex$,\\
such that for $f\in D^{\si\fun\ta}$, $d\in D^\si$: $f d \in D^\ta$,\\
and such that every closed term $M\typ \si$ has its denotation in $D^\si$: 
$\down \set{\eclass{\FPsi M}}{i\geq 0} \in D^\si$.\\
The lubs \wrt $\lex$ will be written $\lubex$ and $\Lubex$, the glbs $\glbex$ and $\Glbex$.
\end{defi}

All f-models coincide on their part of the finite elements \wrt both extensional $\lex$ and stable $\lst$
order.
In the following sections, propositions will mostly deal with finite elements.
The propositions are valid for all f-models if not otherwise stated.

To every f-model we can associate the \defin{semantic map} $\sem{\phantom{M}}\typ \PCF^\si \fun \ENV \fun \Ds$,
where $\ENV$ is the set of environments $\ro$ that map every variable $x^\si$
to some $\ro(x^\si)\in \Ds$.
If $M\typ \si$ is a term with the free variables $x_1,\dots, x_n$, then
\[ \sem{M}\ro = \down \set{\eclass{\FPsi M[x_1:=N_1,\dots,x_n:=N_n]}}{i\geq 0, \eclass{N_j}\in \ro(x_j)}.\]
For closed terms $M$ we also write $\sem{M}$ for $\sem{M}\bo$.

There are three outstanding examples of f-models:
There is the least f-model that consists of just the ideals denoting closed PCF-terms.
There is the greatest f-model consisting of all ideals; this is Milner's and Berry's cpo-model.
And there is the game model consisting of all denotations of (infinite) PCF-B\"ohm-trees,
\ie the sequential functionals.
By Normann's result \cite{Normann} we know that the game model is properly between the
least and the greatest f-models.

Now we will collect the most important properties of f-models.
In the following the $\Ds$ are the domains of some f-model.

\begin{lem}\label{l:finites}
Every $\finsi$ has finitely many elements.\\
The semantics of the $\inf^\si$-terms are the glb-functions with respect to the order $\lex$;
we write $\glbex$ for these functions.\\
If $d,e\in \finsi$, then $d\glbex e \in \finsi$.\\
If $d,e \in \finsi$ are compatible (bounded), \ie there is some $a\in \Ds$ with $d\lex a$ and $e\lex a$,
then there is a lub $d\lubex e\in \finsi$. 
\end{lem}

With this  lemma we can prove:

\begin{prop}
All $\Dst$ are order-extensional, \ie:
\begin{align*}
\text{If } f,g\in \Dst, \text{ then } f\lex g &\ifff \all{d\in \Ds} f d \lex g d\\
                                            &\ifff \all{d\in \fins} f d \lex g d
\end{align*}
\end{prop}

Elements of $\Dst$ will be identified with the corresponding functions.
$\apply$ and these functions are all monotone.
They are continuous with respect to certain directed lubs, the f-lubs.

\begin{defi}
The directed set $S\sub \Ds$ \defin{has the f-lub} $s \in \Ds$, written $S\flub s$,
iff $s$ is an upper bound of $S$ and for all finite $x\lex s$ there is some $y\in S$ with $x\lex y$.
(This is equivalent to: $s$ is the set-theoretical union of $S$.
$s$ is also the lub of $S$ \wrt $\lex$.)\\
A function $f\typ \Ds \fun \Dt$ is \defin{f-continuous}, iff it is monotone and respects f-lubs 
of directed sets $S\sub \Ds$,
\ie if $S\flub s$, then $f S \flub f s$. (With $f S = \set{f x}{x\in S}$.)
\end{defi}

\begin{prop}
The $\apply$ operation is f-continuous on the domain $\Dst\times\Ds$.
(With component-wise order and pairs of finite elements as finite elements.)
Therefore $\apply$ is f-continuous in each argument, and the functions of $\Dst$ are f-continuous.
\end{prop}

In \cite{Normann/Sazonov} it is shown that in the game model there are lubs of directed sets that are not f-lubs;
and that there are finite elements that are not compact in the usual sense with respect to 
general directed lubs.

The f-lubs are exactly the directed lubs for which all functions are continuous:
If we have a directed lub that is not an f-lub,
then this lub contains a finite element that is not contained in the directed set.
The PCF-function that ``observes'' (or ``tests'') this finite element is a function that is not
continuous for the directed set.

In the greatest f-model all lubs of directed sets are f-lubs.
If $S\flub s$ in the greatest f-model,
then the same holds in all f-models that contain $s$ and the elements of $S$.

In an f-model we can define natural lubs in the sense of Sazonov as hereditarily pointwise lubs.
Then a directed set $S$ has the f-lub $s$ iff $S$ has the natural lub $s$.

Side remark: Here we must also mention the ``rational chains'' of Escard\'o and Ho
\cite{Escardo/Ho}.
These are ascending sequences of PCF terms that can be defined syntactically by
a PCF procedure.
The denotations (in any f-model) of the elements of a rational chain always form a directed set with an f-lub (natural lub).
The converse does not hold generally.

\begin{prop}
The semantic map of an f-model fulfills the usual equations, \ie the constants have
their intended meanings, and:
\begin{align*}
\sem{\la x. M}\ro d  &= \sem{M} \ro[x:=d] \\
\sem{M N}\ro  &= \sem{M}\ro \sem{N}\ro \\
\sem{\Y M}\ro &= \Lubex_{n\geq 1} ((\sem{M}\ro)^n \bo)
\end{align*}
\end{prop}

\begin{prop}[Berry, 3.6.11 in \Berry]\label{p:projection}
Define the functions $\fpsi = \sem{\FPsi}\bo \typ \Ds \fun \Ds$.\\
For all $\si$, $(\fpsi)$ is an increasing sequence of finite projections with f-lub the identity $\id$:
\begin{align*}
\fpsi  &\lex \id \\
\fpsi \comp \fpsi  &= \fpsi, \text{ with $\comp$ function composition} \\
\fpsi  &\lex \fp^\si_{i+1} \\
\set{ \fpsi }{i\geq 0}
 &\flub \id  \\
\fpsi(\Ds) &= \finsi
\end{align*}
\end{prop}

\begin{prop}
Every f-model is fully abstract for PCF: For all terms $M$, $N$ of the same type
\[(\all{\ro \in \ENV} \sem{M}\ro \lex \sem{N}\ro)  \ifff  M \lop N .\]
\end{prop}

\bigskip
In the rest of this section we will define the stable order in f-models
and collect the corresponding properties that will be needed in this paper.

The definition of the stable order $\lst$ is given by Berry \cite[4.8.6, page 4-93]{Berry}
for the fully abstract cpo-model as follows:
\begin{align*}
\text{For } d,e \in D^\io : d\lst e  \ifff {} & d\lex e \\
\text{For } f,g \in \Dst :  f\lst g  \ifff {} & \all{x\in \Ds} f x \lst g x \text{ and}\\
  & \all{x,y \in \Ds } x\cost y  \ifthen  f x \glbex g y = f y \glbex g x
\end{align*}
(Here $\cost$ means compatibility \wrt $\lst$.)\\
This definition serves as well for our f-models, but I prefer the equivalent (\wrt the full type hierarchy) form:

\begin{defi}[stable order $\lst$]
\begin{align*}
\text{For } d,e \in D^\io : d\lst e  \ifff {} & d\lex e \\
\text{For } f,g \in \Dst :  f\lst g  \ifff {} 
  & \all{x,y \in \fins} x\lst y  \ifthen  f x = f y \glbex g x  \db \\
\intertext{The order $\lst$ is extended pointwise to environments from $\ENV$,
here used in the definition of $\lst$ on denotations:}
\text{For } f,g \in \ENV \fun \Ds : f\lst g \ifff {} 
  & \all{\ro,\varepsilon \in \ENV} \ro \lst \varepsilon \ifthen f\ro = f\varepsilon \glbex g \ro 
\end{align*}
The lubs \wrt $\lst$ will be written $\lubst$ and $\Lubst$, the glbs $\glbst$.
\end{defi}

Note that $\glbex$ is by definition the glb \wrt the \emph{extensional} order $\lex$.
But we can prove the following:

\begin{prop}
In any actual f-model the following holds:\\
For $f,g \in \Ds$:
If $f,g$ are $\lst$-compatible in the greatest f-model,
then $f\glbex g$ is also the glb \wrt $\lst$.
(Note: If $f,g$ are $\lst$-compatible in the actual f-model,
then they are also compatible in the greatest f-model.)\\
If $f\lst g$ then $f\lex g$.
$\lst$ is a partial order on $\Ds$.
\begin{align*}
\text{For } f,g \in \Dst :  f\lst g  \ifff {} & \all{x\in \Ds} f x \lst g x \text{ and}\\
  & \all{x,y \in \Ds } x\lst y  \ifthen  f x = f y \glbex g x  \db \\
\intertext{The definition of $\lst$ can be given in ``uncurried'' form with vectors of arguments, the order $\lst$
      extended componentwise:}
\text{For } f,g \in \Dsn : f\lst g \ifff {} &
     \all{x_1,y_1\in D^{\si_1}, \ldots, x_n,y_n\in D^{\si_n}}\\
    & (x_1,\ldots, x_n) \lst (y_1, \ldots, y_n)
          \ifthen\\
    &  fx_1 \ldots x_n = f y_1 \ldots y_n \glbex g x_1 \ldots x_n
\end{align*}
\end{prop}
\proof
The proof that $f\glbex g$ is the glb \wrt $\lst$ (for $\lst$-compatible $f,g$)
is by induction on the type $\si$.
It uses only the definition of $\lst$ and that $\glbex$ is the glb \wrt $\lex$,
no stability (or conditional multiplicativity) is used.
\qed

\begin{defi}
$f\in \Dst$ is \defin{conditionally multiplicative (cm)} if
\[ \all{x,y\in \fins} 
                 x \cost y  \ifthen  f(x \glbex y) = f x \glbex f y \]
Analogously for denotations $f\in \ENV \fun \Ds$.
\end{defi}

This definition can also be given in ``uncurried'' form:
$f \in \Dsn$ is cm iff
\begin{align*}
   \all{x_1,y_1\in D^{\si_1}, \ldots, x_n,y_n\in D^{\si_n}}
   & (x_1,\ldots, x_n) \cost (y_1, \ldots, y_n)
          \ifthen \\
  & f(x_1 \glbex y_1) \ldots (x_n \glbex y_n) =  f x_1 \ldots x_n \glbex f y_1 \ldots y_n  
\end{align*}

\begin{thm}[Berry, 4.8.10 in \Berry]
In an f-model, all functions from domains $\Dst$ are cm.
All denotations $\sem{M}$ are cm.
\end{thm}
\proof
Berry first proves the property cm for the denotations of normal form terms by induction on the size of the type.
Then it is extended to all functions by continuity.
\qed

\begin{prop}[Berry \Berry, syntactic monotony \wrt $\lst$] \hfill { } \\
For every context $C[{\phantom{M}}]$ with hole of type $\si$, and terms $M,N \typ \si$:\\
If $\sem{M} \lst \sem{N}$ then $\sem{C[M]} \lst \sem{C[N]}$.\\
Therefore, for terms $M,N\typ \si$: If $M\lsy N$ then $\sem{M} \lst \sem{N}$.
\end{prop}

We will also write $M\lst N$ for $\sem{M} \lst \sem{N}$.

Now we show property I of $(\Ds,\lst)$ and the representation of all functions by traces,
which is not contained in Berry's thesis.

\begin{prop}\label{p:propI}
For the finite projections we have: $\fpsi \lst \fp^\si_{i+1}$ and $\fpsi \lst \id$.\\
The $\finsi$ are downward closed \wrt $\lst$:
If $d\in \finsi$, $e \in \Ds$ and $e\lst d$, then $e \in \finsi$.\\
Therefore the domains $(\Ds,\lst)$ have the property I: There are only finitely many elements under each
finite element.
\end{prop}
\proof
The proof of $\fpsi \lst \id$ is by induction on the type $\si$;
the induction step is in the proof of proposition 12.4.4 in the section on stable bifinite
domains of \cite[page 287]{Amadio/Curien}.
The downward closedness of $\finsi$ is an easy consequence and can be found at the same place.
\qed

Because of property I, all our functions of $\Dst$ (which are cm) are also stable, 
and therefore can be represented by traces.
We chose the trace of the uncurried form.

\begin{defi}
Let $f\in \Dsn$, $n\geq 0$, $x_i \in D^{\si_i}$ and $f x_1 \ldots x_n = j$ for some integer $j$.\\
Then there are $y_i\in \fin^{\si_i}$, $y_i \lst x_i$, with $f y_1 \ldots y_n = j$
and $(y_1,\ldots,y_n)$ is the $\lst$-least vector with this property.
(This is the meaning of: $f$ is stable.)\\
In this case we say that $y_1 \mto \ldots \mto y_n \mto j$ is a \defin{token} of $f$.\\
The set of all tokens of $f$ is called the \defin{trace} of $f$, written $\trace(f)$.\\
The $y_i$ in the token will be represented by traces again.
We will use a liberal syntax for tokens and traces,
writing $\bo$ for the trace $\emptyset$,
$\0$ for the trace $\{\0\}$ of $\0$,
and also $\0\0\mto \0$ for the token $\{\0\}\mto\{\0\}\mto\0$.
If $M$ is a closed term, we write simply $\tsem{M}$ for the trace of its denotation $\sem{M}\bo$.
\end{defi}

\begin{prop}
For $f,g\in \Ds$: $f\lst g$ iff $\trace(f) \sub \trace(g)$.\\
If $f,g$ are $\lst$-compatible in the greatest f-model,
then $\trace(f\glbex g) = \trace(f) \cap \trace(g)$.\\
$f\in \Ds$ is finite of grade $i$, $f\in \finsi$, iff all numbers in the trace of $f$ are $\leq i$.
\end{prop}

\section{Game Terms}

\newcommand{\G}%
{\tlf{\la f g.}{g}{\longarm{\tf{g}{\tz\tbo}{\ti}{\tbo}}{\treesp{-7mm}\tbo}}%
{\longarm{\tf{f}{\ttlx{x}{x}{\tz\ti}}{\ttwo}{\ttwo}}}%
{\treesp{-10mm}\tbo}}

Berry's conjectures demand the existence of certain finite PCF-terms.
In this section we show that we may restrict these finite terms to terms
in a certain standard normal form that we call \emph{game terms}.
This will simplify the proofs of the counter-examples, and is also an interesting result itself.
Game terms first appeared in the literature on game semantics as terms
representing game strategies;
in \cite[section 3.2]{Abramsky/Jagadeesan} they were called (finite and infinite) ``evaluation trees'',
in \cite[section 7.3]{Hyland/Ong} ``finite canonical forms'' that correspond to compact innocent strategies,
and in \cite[section 6.6]{Amadio/Curien} ``PCF B\"ohm trees''.
The textbook article on ``PCF B\"ohm trees'' comes closest to our approach,
as it introduces a semantics in the form of B\"ohm trees and has to solve similar problems
in the needed syntactic transformations.
But we do not employ a (game or other) semantics, \ie we do not interpret the PCF-constants
by infinite strategies or B\"ohm trees; our approach is purely syntactic.
We take a finite PCF-term, apply an operator that resembles the finite projection $\FPsi$
and reduce the resulting term to its game term form.
We show that the transforming reductions respect the syntactic order $\lsy$
(used in the refutation of Berry's second conjecture),
and this will also enable us to proceed to infinite game terms.
We also introduce a graphical representation of game terms that makes the behaviour of
terms better visible.

\bigskip

First we introduce an additional new construct for the PCF language, for every $i\geq 0$:\\
If $M, N_0, \ldots, N_i \typ \io$, then $\casei M \Ni \typ \io$.\\
Please note that $\casei$ is not a constant, but the whole case-expression is a
new construct of the language, it is no application.
We call the new terms (PCF-)case-terms,
and a case-term with all case-expressions as $\casei$ for fixed $i$ we call $\casei$-term.
The reduction rule for $\casei$ is:
\[ \casei n \Ni \re N_n, \text{ for } 0\leq n \leq i\]
The case-expression is equivalent to a PCF-term:
\[ \casei M \Ni \eq \pif M \pthen N_0 \pelse \pif \ppre^1 M \pthen N_1 \pelse 
   \ldots \pif \ppre^i M \pthen N_i \pelse \Om \]
This is the ``filter'' as it appears in the finite projection term $\FP^\io_i$.
So $\casei$ does not enhance the expressiveness of PCF.
It is merely a ``macro'' that is used as short expression for the filter term above,
to keep the unity of the filter term in the transformation to game terms.

The syntactic order $\lsy$ is defined on case-terms as follows:
\[ \casei M \Ni \lsy \casej M' N'_0 \ldots N'_j \text{ iff } i\leq j,\; M\lsy M'
\text{ and } N_k\lsy N'_k \text{ for } 0\leq k\leq i. \]
This is equivalent to the syntactic order on the macro expansions of the case-expressions.

\begin{defi}\label{d:gameterm}
\defin{Game terms} are the well-typed PCF-case-terms that
are furthermore produced by the following grammar:
\begin{align*}
M,N ::= {} & \Om^\si, \text{ $\si$ any type} \\
        & \laxn m, \text{ $m$ integer constant, $n \geq 0$} \\
        & \laxn \casei (y \Mm ) \Ni, \text{ $y$ variable, $n,m,i \geq 0$}
\end{align*}
Please note that $\laxn$ vanishes for $n=0$, so needed for the $N_k$ of type $\io$.

A \defin{game term of grade $i$}, $i\geq 0$, is a game term that is a $\casei$-term (every $\case$ is $\casei$)
with all integer constants $\leq i$.
(This entails that a closed game term of grade $i$ is a finite term of grade $i$.)



A \defin{game term of pregrade} $i$, $i\geq 0$, is a game term that is furthermore
produced by the following grammar for the non-terminal $N$:
\begin{align*}
N ::= {} & \Om^\si, \text{ $\si$ any type} \\
        & \laxn m, \text{ $m$ integer constant, $n \geq 0$} \\
        & \laxn \casei (y \Mm ) \Ni, \text{ $y$ variable, all $M_k$ game term of grade $i$,}\\
        &\phantom{\laxn \casei (y \Mm ) \Ni,} n,m \geq 0
\end{align*}
(A game term of pregrade $i$ is a $\casei$-term.)
\end{defi}

Informally, we call the positions in a game term of integer constants at the top level,
\ie where this integer serves as output of the term, \defin{output positions}.
So a game term of pregrade $i$ is a game term such that for all integer constants $m$ that are
\emph{not} in output position it is $m\leq i$.
(So the integers at output positions are not restricted.) 

We define a notion for the replacement of integers in output positions of game terms.
\begin{defi}
Let $P,L$ be game terms, $L\typ \io$ and $l\geq 0$.
We define $P\outrep{l:=L}$ by recursion on $P$:
\begin{align*}
\bo\outrep{l:=L} & = \bo \\
(\laxn l)\outrep{l:=L} &= \laxn L \\
(\laxn m)\outrep{l:=L} &= \laxn m, \text{ for } m\neq l \\
(\laxn \casei (y \Mm) \Ni)\outrep{l:=L} &= \laxn \casei(y\Mm) \\
  &\phantom{= \laxn \casei} N_0\outrep{l:=L}\ldots N_i\outrep{l:=L}
\end{align*}
We also write multiple replacements, \eg $P\outrep{l:=L_l \text{ for } l\geq 0}$.
These multiple replacements are done in parallel, the whole replacement moves down the term.
\end{defi}


We will use a graphical representation of game terms in the next sections:\\
A subterm 
        $\laxn \casei (y \Mm ) \Ni$ is represented in the graph by a node of the form:

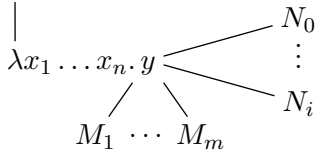
\pstree[treemode=D]{\TR{ }}
{\ttlf[20mm]{x_1 \ldots x_n}{y}
                    {\TR{$M_1$}\Tn\rput(0,0){$\ldots$}\TR{$M_m$}}
                    {\TR{$N_0$}\Tn\rput(0,-4mm){$\vdots$}\TR{$N_i$}}}

\bigskip

The upper parent of this node is connected to the $\la$;
if the $\la$ is missing, the upper or left parent is connected to the $y$.
The $M_1,\ldots, M_m$ are the \emph{legs} of $y$;
the $N_0, \ldots, N_i$ are the \emph{arms} of $y$.
A leg or arm that points to a $\Om$ is mostly represented simply by a leg or arm pointing to empty space.
This graphical representation makes the behaviour of game terms much better visible.

\medskip
\no
\textbf{Example:}

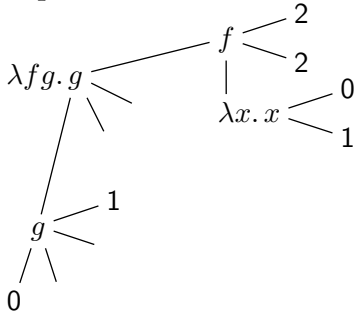
\nopagebreak

\G

\bigskip
This is the representation of the term:
\[ \la f g. \caseum[g(\caseum(g \0 \Om)\1\Om)\Om][\caseum(f(\la x.\caseum x \0\1))\2\2]\Om \]
of type $((\io\fun\io)\fun\io)\fun(\io\fun\io\fun\io)\fun\io$.
It is a game term of \emph{pregrade} $1$.
The output positions are the two positions of the number $\2$.
If we replace the number $\2$ at the output positions by $\Om$, $\0$ or $\1$,
then we get a game term of \emph{grade} $1$.

Game terms are the real ``medium'' in which to investigate Berry's problems:
First, if one seeks terms $M$ which have many semantically different syntactic parts $N\lsy M$,
according to Berry's second conjecture, then one is naturally led to game terms,
because they have a very fine syntactic structure.
Second, they simplify the proofs of the counter-examples.
The conditional always appears together with a variable,
cutting down the cases to be analysed and simplifying the induction hypotheses considerably.

In the next subsection we develop a map $\gti$ from finite terms to equivalent game terms
such that $M\lsy N \typ \si$ entails $\gti(M) \lsy \gtj(N)$, where $M,N$ are of grade $i$ \resp $j$, $i\leq j$.
This means that the refutation of Berry's conjectures may be restricted to game terms.
In the following subsection we extend our result to infinite game terms.
They are needed for a full formulation of Berry's conjectures for first-order types (where they are valid).

\subsection{Finite Game Term Theorem}

We are given finite terms $M\lsy N$ and want to find equivalent game terms.
First we must get rid of the $\Y$s in the terms.

The map $\om\typ \PCF^\si \fun \PCF^\si$ (for all types $\si$) is taken from \cite{Berry/Curien,Berry}
and called the \defin{immediate syntactic value}:
\begin{equation*}
\om(M) = 
\begin{cases}
 \laxn u\, \om(M_1) \ldots \om(M_m),
           & \text{if } M=\laxn u \Mm\\
           & \text{with $u$ a variable or constant,}\\
           & \text{\ie $M$ is in head normal form} \\
         \Om \text{ else}
\end{cases}
\end{equation*}
Please note here that a constant is $\psuc, \ppre, \pif, \0, \1, \2, \ldots$
A constant is not $\Om$ or $\Y$.\\
$\reby$ is the one-step reduction with the $\beta$-rule or the rule $\Y M \re M(\Y M)$ in any context.\\
As is known from \cite{Berry/Curien,Berry}, if $M\rebyt N$, then $\om(M) \lsy \om(N)$.

\begin{lem}[Approximation Lemma]\label{l:approx1}
For every finite term $M$ there is a term $N'$ such that $M\rebyt N$ for some $N$,
$N'\lsy \om(N)$, $M\eq N'$ and $N'$ is the $\lsy$-least term with this property.
This unique $N'$ is called $\approx(M)$.
\end{lem}
\proof
For the fully abstract cpo-model (and therefore for all f-models) the approximation continuity theorem
\cite[theorem 4.3.1]{Berry/Curien}
is valid:
\[ \set{\sem{\om(N)}} {M\rebyt N} \flub \sem{M}. \]
The set on the left is directed and $M$ is finite,
therefore there is $N$ with $M\rebyt N$ and $\sem{M} = \sem{\om(N)}$.

Now assume the type of $M,N$ is $\sinio$.
Take any vector of closed terms $A_1\typ \si_1,\ldots , A_n\typ \si_n$
with $\om(N) \An \ret m$ (integer constant).

By syntactic stability \cite[theorem 2.8.8]{Berry} \cite[theorem 3.6.7]{Berry/Curien}
there is a $\lsy$-least term $\Ns\lsy \om(N)$ with $\Ns \An\ret m$.
Take as $N'$ the $\lsy$-lub of all these $\Ns$.
\qed

\begin{lem}\label{l:approx2}
For all finite terms $M\lsy N$ it is $\approx(M)\lsy \approx(N)$.
\end{lem}
\proof
Let $M'$ be a term with $M\rebyt M'$ and $M\eq \om(M')$.\\
As the $\beta$-rule and the $\Y$-rule do not involve $\Om$,
all these reductions $M\rebyt M'$ can also be done in $N$.
(If $A\lsy B$ and $A\reby A'$, then there is $B'$ with $B\reby B'$ and $A'\lsy B'$.)\\
So there is $N'$ with $N\rebyt N'$ and $M'\lsy N'$,
and of course $\om(M') \lsy \om(N')$.

By confluence of $\reby$ there is $N''$ with $N'\rebyt N''$ and $N\eq \om(N'')$.\\
It is $\om(N') \lsy \om(N'')$, therefore $\om(M')\lsy \om(N'')$.\\
$\approx(M)$ is the least term $X$ with $X\lsy \om(N'')$ and $M\lop X$.\\
$\approx(N)$ fulfills the two conditions for $X$, therefore $\approx(M)\lsy \approx(N)$.
\qed

Now we have finite terms $\approx(M)\lsy \approx(N)$ without $\Y$.
The next step is to apply a $\FPsi$-like operator to the terms and reduce according to some reduction rules
to game terms.
The proof can be done in different ways:

In my first version I proved the termination of the reductions, formulated an invariant of 
the (eta-expanded) term structure,
proved the invariance under the reductions and that they lead to game terms.
This resulted in an induction on the reduction sequence,
the induction step done by induction on the term, causing much rewriting bureaucracy.
(This ugly proof is available as supplementary material from my home page.)

Here we will see a more elegant half-sized proof based on an induction on the term from the beginning,
with the aid of a reducibility predicate (see \eg \cite[theorem 3.1]{Plotkin}).
(Jim Laird also uses a reducibility predicate to produce eta-expanded normal forms of a simply typed
$\la$-calculus with lifting (without inconsistent values) \cite[proposition 4.2]{Laird:sequ}.)

To produce the game terms we define for every $i\geq 0$ a big-step reduction relation $M\redi N$ on $\casei$-terms.
The mere existence of the game terms could be proved without $\redi$,
but we want to give an explicit deterministic algorithm.
(Determinism is easily built into big-step reduction.)
The values for $\redi$, \ie the terms that we consider as the results of reductions,
are the game terms of pregrade $i$.



\newcommand{\xskip}{\bigskip\no}
\newcommand{\infsp}{\;\;}


\xskip
Here are the rules for $\redi$.
In the hypothesis of a rule the abbreviation $M\redi N \; gi$ means
``$M\redi N$ and $N$ is a game term of grade $i$'',
$M\redi N\; pi$ means
``$M\redi N$ and $N$ is a game term of pregrade $i$''.

\xskip
(0)\infsp $n\redi n$ for all integer constants $n$

\xskip
(1) \infsp \infer{M[y:=M_1]M_2\ldots M_m \redi P}{(\la y.M)M_1 M_2 \ldots M_m \redi P}, for $m\geq 1$
\hspace{10mm} 
(2) \infsp $\bo \Mm \redi \bo$, for $m\geq 0$

\xskip
(3)\infsp \infer{A\redi A'\; pi}{\psuc A \redi A'\outrep{m:=m+1\text{ for }m\geq \0}}
\hfill       
(4)\infsp \infer{A\redi A'\; pi}{\ppre A \redi A'\outrep{\0:=\bo,\; m:=m-1\text{ for }m\geq \1}}

\xskip
(5)\infsp \infer{A_k\redi A'_k \; pi,\text{ for } k=1,2,3}
                {\pif A_1 \pthen A_2 \pelse A_3 \redi A'_1\outrep{\0:=A'_2,\; m:=A'_3 \text{ for } m\geq 1}}

\xskip
(6)\infsp \infer{A\redi A'\; pi, A'\neq\bo}
                {\laxn \casei A \0 \ldots i \redi \laxn A'\outrep{k:=\bo \text{ for } k>i}}, for $n\geq 0$

\xskip
(7)\infsp \infer{A\redi \bo}{\laxn \casei A \0\ldots i \redi \bo}, for $n\geq 0$

\xskip
(8)\infsp \infer{A_k\redi A'_k\; gi,\;\text{ for } 1\leq k\leq m}
                {\casei(x A_1\ldots A_m) \0 \ldots i \redi \casei(x A'_1\ldots A'_m)\0\ldots i}, for $m\geq 0$

\bigskip
Remarks: Not for all $\casei$-terms $M\typ \si$ there is a value $V$ with $M\redi V$,
but there will be a value $V$ with $\FPsi M \redi V$ for $\FPsi$ suitably defined.
The reduction relations are complete enough for the purposes of the following proofs.
So to understand the reductions at this stage,
just check the soundness of each rule separately, according to the following lemma,
and do not bother about completeness.
When you go through the subsequent proofs, you will see that exactly these rules are needed,
no more, no less.

\begin{lem}[soundness of the reduction relations $\redi$]
For all $\casei$-terms $M,M'$:
If $M\redi M'$, then $\sem{M} = \sem{M'}$ and $M'$ is a value (\ie a game term of pregrade $i$).
\end{lem}
\proof
Translate each reduction rule into a rule with semantic equivalence instead of the reduction relation:
Translate statements $A\redi A'$ into ($\sem{A}=\sem{A'}$ and $A'$ is a value),
and keep the statements $gi$ and $pi$.
Then check each translated rule for validity.
\qed

Now we come to the reducibility predicate.
We pack all that we want to prove into its definition:
the compatibility of the transformation with the order $\lsy$ and even the uniqueness of the reduction $\redi$.

\begin{defi}[reducibility predicate]
Let $i\leq j$, $A$ a $\casei$-term and $B$ a $\casej$-term of type $\si=\sinio$, $n\geq 0$.\\
$A\lsy B \typ \si$ are $(i,j)$-\defin{transformable}, written $A\lsy B\typ\si(i,j)$,\\ iff
for all $A_l\lsy B_l\typ \si_l(i,j)$, $1\leq l\leq n$,
there are game terms $A',B'\typ \io$ of pregrade $i$ \resp $j$ with
$A\An \redi A'$ and $B\Bn\redj B'$,\\
$A'$ and $B'$ are unique for these reductions, and furthermore $A'\lsy B'$.
\end{defi}

Note that this definition does not take care of the free variables of $A,B$.
Note also that it does not demand the \emph{grade} $i,j$ of $A',B'$, but the \emph{pregrade}.
So it will be applicable to general terms that do not restrict the integer constants, in lemma
\ref{l:transformable}.

\begin{lem}\label{l:bottom}
If $A\lsy B\transi$, then
$\bo\lsy B \typ \si(k,j)$ for all $k\leq j$.
\end{lem}
\proof
Easy consequence of the definition of the reducibility predicate and
of rule (2) for $\bo$-application.
\qed 

For the next lemma we need a notion of simultaneous substitution for PCF-terms that properly
renames bound variables.
We take Allen Stoughton's definitions \cite{Stoughton:substitution}.

A \defin{substitution} is a function $s,t$ from variables to terms (of the type of the variable).
The substitution $s[x:=N]$ is defined by $(s[x:=N])x = N$
and $(s[x:=N])y = sy$ for $y\neq x$.
$\id$ is the identity substitution.

If $x$ is a variable, $M$ a term, $s$ a substitution, then we define
\[ \new xMs = \set{y}{y \text{ variable and for all } z \in FV(M)-\{x\}.\; y \not\in FV(sz)}, \]
where $FV(X)$ is the set of free variables of term $X$.

The \defin{simultaneous substitution} $Ms$ of $sx$ for the free occurrences of $x$ in $M$,
for all $x$, is defined by structural recursion on $M$:
\begin{align*}
xs &= sx, \text{ for every variable $x$} \\
cs &= c, \text{ for every constant $c$} \\
(MN)s &= (Ms)(Ns) \\
(\la x.M)s &= \la y.(M(s[x:=y])), \text{ with } y=\choice(\new xMs),
\end{align*}
where $\choice$ is a fixed function that chooses some variable $y$ from the argument set of variables.

We suppose that the normal substitution (in the $\beta$-rule) behaves like this:
\[ P[y:=N] = P(\id[y:=N]). \]

\begin{lem}\label{l:subst}
For terms $M,N$, substitution $s$ and variables $x,y$ with $y=\choice(\new xMs)$ we have:
\[ (M(s[x:=y]))[y:=N] = M(s[x:=N]) \]
\end{lem}
\proof
Follows from theorem 3.2 of \cite{Stoughton:substitution}.
\qed

\begin{lem}\label{l:transformable}
Let $A\lsy B\typ \si$ be PCF-terms without $\Y$.\\ 
Let $\{x^{\ta_1}_1,\ldots,x^{\ta_m}_m\}$
be a superset of the free variables of $B$.\\
For $1\leq k\leq m$ let $A'_k\lsy B'_k\typ \ta_k(i,j)$ be $\casei$- \resp $\casej$-terms that
are $(i,j)$-transformable.\\
Define the substitutions $s=\id[x_1:=A'_1]\ldots [x_m:=A'_m]$
and $t=\id[x_1:=B'_1]\ldots [x_m:=B'_m]$.\\
Then $As\lsy Bt\transi$.
\end{lem}

\proof By induction on the term $B$.
(Note: PCF-terms are without $\case$.)

\no
\textbf{Case} $B=\Bs B_0$, $\Bs \typ \si_0 \fun \sinio$, for $n\geq 0$:\\
First let $A=\As A_0$.\\
By the induction hypothesis we get $\As s \lsy \Bs t\typ \si_0\fun\ldots \si_n\fun \io(i,j)$
and $A_0 s \lsy B_0 t\typ \si_0(i,j)$.\\
Let $A_l\lsy B_l\typ \si_l(i,j)$ for $1\leq l \leq n$.\\
By the reducibility predicate there are game terms $A'\lsy B'\typ\io$ of pregrade $i$ \resp $j$ with
\begin{align*}
(\As s)(A_0 s) \An &\redi A' \\
(\Bs t)(B_0 t) \Bn &\redj B' 
\end{align*}
So $As \lsy Bt \transi$.

Now let $A=\Om$.
By the same argument we have $Bt\lsy Bt\typ \si(j,j)$,
therefore by lemma \ref{l:bottom}: $\bo\lsy Bt\transi$.

\medskip
\no
\textbf{Case} $B=\la x.\Bs\typ \sinio$, $n\geq 1$:\\
First let $A=\la x.\As$.\\
Let $A_l\lsy B_l\typ \si_l(i,j)$ for $1\leq l\leq n$.\\
By the induction hypothesis for $\Bs$ we get 
\[ \As(s[x:=A_1])\lsy \Bs(t[x:=B_1]) \typ \si_2\fun \ldots \fun \si_n\fun \io (i,j). \]
Therefore there are game terms $A',B'\typ \io$  of pregrade $i$ \resp $j$ with
\begin{align*}
(\As(s[x:=A_1]))A_2\ldots A_n &\redi A'\\
(\Bs(t[x:=B_1]))B_2\ldots B_n &\redj B',
\end{align*}
with $A',B'$ unique and $A'\lsy B'$.\\
By lemma \ref{l:subst} and the definition of substitution we get: 
\begin{align*}
\As(s[x:=A_1]) &= (\As(s[x:=y]))[y:=A_1], \text{ for } y=\choice(\new x\As s) \\
\Bs(t[x:=B_1]) &= (\Bs(t[x:=z]))[z:=B_1], \text{ for } z=\choice(\new x\Bs t) \\
(\la x. \As)s &= \la y. \As(s[x:=y]) \\
(\la x. \Bs)t &= \la z. \Bs(t[x:=z]) 
\end{align*}
Then it reduces
\begin{align*}
(\As(s[x:=y]))[y:=A_1] A_2 \ldots A_n &\redi A', \text{ and therefore by rule (1):} \\
(\la y. \As(s[x:=y])) A_1 A_2 \ldots A_n &\redi A', \text{ therefore} \\
(\la x. \As)s A_1 A_2 \ldots A_n &\redi A'.\\
\intertext{Analogously:}
(\la x. \Bs)t B_1 B_2 \ldots B_n &\redj B'.
\end{align*}
These reductions are unique, and $A'\lsy B'$.
So $As \lsy Bt \transi$. \\

Now let $A=\Om$.
By the same argument we have $Bt\lsy Bt\typ \si(j,j)$,
therefore by lemma \ref{l:bottom}: $\bo\lsy Bt\transi$.

\medskip
\no
\textbf{Cases} $B=x$ (variable), $B=n$ (integer constant), $B=\Om$ are clear.\\
For $B=n$ rule (0) is used, for $B=\bo$ rule (2).\\  
For the subcases $A=\bo$ lemma \ref{l:bottom} is used.

\medskip
\no
\textbf{Case} $B=\pif$:\\
First let $A=\pif$.\\
Let $A_l\lsy B_l\typ \io(i,j)$ for $1\leq l\leq 3$.\\
Then there are $A_l\redi A'_l$ and $B_l\redj B'_l$ ($A'_l,B'_l$ unique) with $A'_l\lsy B'_l$,
for $1\leq l\leq 3$.\\
It reduces by rule (5):
\begin{align*}
\pif A_1 \pthen A_2 \pelse A_3 &\redi A'_1\outrep{\0:=A'_2,\; m:= A'_3 \text{ for }m\geq 1}\\
\pif B_1 \pthen B_2 \pelse B_3 &\redj B'_1\outrep{\0:=B'_2,\; m:= B'_3 \text{ for }m\geq 1} 
\end{align*}
Both reductions are unique and the results are in relation $\lsy$.

\no Now let $A=\Om$. 
By lemma \ref{l:bottom} it is $\bo\lsy \pif \transi$.

\medskip
\no
\textbf{Cases} $B=\psuc$, $B=\ppre$: analogous to $B=\pif$.\\
For $B=\psuc$ rule (3) is used, for $B=\ppre$ rule (4).
\qed

Next we prove a lemma that introduces the terms $\FPsi$ into the transformation.
For the rest of this section we redefine the finite projection terms $\FPsi$ as equivalent $\casei$-terms:
\[ \FP^{\sinio}_i = \la f. \laxn \casei[f \FPixn] \0 \ldots i, \text{ for } n\geq 0. \]

\begin{lem}\label{l:projection}
For all types $\si=\sinio$ the following three propositions are valid:
\begin{enumerate}[(1)]
\item For all $A\lsy B\transi$ it is 
\[ A(\FP^{\si_1}_i x_1)\ldots (\FP^{\si_n}_i x_n)
\lsy B(\FP^{\si_1}_j x_1)\ldots (\FP^{\si_n}_j x_n) \typ \io(i,j).\]
\item For all $A\lsy B\transi$ there are $A',B'\typ \si$ with $\FPsi A \redi A'$ and $\FP^\si_j B \redj B'$
such that both are unique for this reduction,
and furthermore $A'\lsy B'$ and they are game terms of \emph{grade} $i$ \resp $j$.
\item For all variables $x^\si$ and $i\leq j$: $\FPsi x^\si \lsy \FP^\si_j x^\si \transi$.
\end{enumerate}
\end{lem}

\proof By simultaneous induction on the type $\si$.

\no
\textbf{(1)}
By the induction hypothesis for (3) we get $\FP^{\si_k}_i x_k \lsy \FP^{\si_k}_j x_k \typ \si_k(i,j)$,
for $1\leq k \leq n$, and the proposition follows.

\no
\textbf{(2)}
The proposition (1) means that there are game terms $A'',B''\typ \io$ with pregrade $i$ \resp $j$ such that
$A \FPixn \redi A''$ and $B \FPjxn \redj B''$,
with $A'',B''$ unique for this reduction and $A''\lsy B''$.\\
If $A''=\bo$ then it reduces by rule (7):
\[ \laxn \casei[A \FPixn ] \0\ldots i \redi \bo \]
and therefore by rule (1): $\FPsi A \redi \bo$.\\
If also $B''=\bo$, then likewise $\FPsj B \redj \bo$ and the proposition follows.\\
(We still have $A''=\bo$.)
If $B''\neq \bo$ then it reduces by rule (6):
\[ \laxn \casej[B \FPjxn ] \0\ldots j \redj \laxn B''\outrep{k:=\bo\text{ for } k>j} = B' \]
and therefore by rule (1): $\FPsj B \redj B'$, $B'$ is a game term of grade $j$,
and the proposition follows.

\no If $A''\neq \bo$ and $B''\neq \bo$,
then we get like the last reduction by rules (6) and (1):
\begin{align*}
\FPsi A &\redi \laxn A''\outrep{k:=\bo\text{ for } k>i} = A'\\ 
\FPsj B &\redj \laxn B''\outrep{k:=\bo\text{ for } k>j} = B' 
\end{align*}
Both reductions are unique, it is $A'\lsy B'$ and they are game terms of grade $i$ \resp $j$.

\no
\textbf{(3)}
We have to prove that for all $A_l \lsy B_l \typ \si_l(i,j)$, $1\leq l \leq n$,
there are game terms $A'\lsy B'$ of pregrade $i$ \resp $j$ with
$(\FPsi x) \An \redi A'$ and
$(\FPsj x) \Bn \redj B'$ (with uniqueness of the reductions).

By the induction hypothesis of (2) for all $l$ there are game terms $A'_l \lsy B'_l\typ \si_l$
of grade $i$ \resp $j$ with
$\FP^{\si_l}_i A_l \redi A'_l$ and
$\FP^{\si_l}_j B_l \redj B'_l$ (with uniqueness of the reductions).\\
It reduces by rule (8)
\[ \casei[x (\FP^{\si_1}_i A_1) \ldots (\FP^{\si_n}_i A_n)] \0\ldots i \redi
 \casei [x A'_1\ldots A'_n] \0 \ldots i = A' \]
and therefore by rule (1):
\[ (\FPsi x) \An \redi A' \]
Likewise it reduces by rules (8) and (1):
\[ (\FPsj x) \Bn  \redj
  \casej [x B'_1\ldots B'_n] \0 \ldots j = B' \]
$A',B'$ are even game terms of \emph{grade} $i$ \resp $j$.
The reductions are unique.
It is $A'\lsy B'$.
\qed

\begin{defi}
Let $A$ be a $\casei$-term without $\Y$ with $A\lsy A\typ\si(i,i)$.\\
The unique game term $A'$ of grade $i$ with $\FPsi A \redi A'$ is called $\projsi(A)$.\\
For every finite term $M\typ\si$ we get $\approx(M)$ without $\Y$ with
$\approx(M)\lsy \approx(M)\typ \si(i,i)$ by lemma \ref{l:transformable}.
(Note that finite terms are closed.)\\
We define the map $\gti(M) = \projsi(\approx(M))$,
for $M\typ \si$ finite term of grade $i$.
\end{defi}

\begin{thm}[Game Term Theorem]\label{t:gameterm}
If $i\leq j$ and $M\lsy N\typ \si$ are finite PCF-terms of grade $i$ \resp $j$, then
$\gti(M) \lsy \gtj(N)$ are game terms of grade $i$ \resp $j$ with
$M\eq \gti(M)$ and $N\eq \gtj(N)$.
\end{thm}
\proof
By lemma \ref{l:approx1} and \ref{l:approx2} we get
$\approx(M)\lsy \approx(N)$ without $\Y$.
By lemma \ref{l:transformable} it is $\approx(M)\lsy \approx(N)\typ \si(i,j)$.
By lemma \ref{l:projection}(2)
$\projsi(\approx(M))\lsy \projsj(\approx(N))$ are game terms of grade $i$ \resp $j$.
Furthermore $M\eq \FPsi(\approx(M)) \eq \projsi(\approx(M))$
and likewise for $N$.
\qed

\subsection{Infinite game terms}

\begin{defi}
An \defin{infinite game term} of type $\si$ is an ideal of game terms of type $\si$ (of any grade),
under the ordering $\lsy$.
(Infinite game terms can be construed as B\"ohm trees with infinite $\case$-expressions,
which we write as $\caseinf M N_0 N_1 \ldots$.)
The order $\lsy$ on infinite game terms is the subset order of the ideals.
The semantics (in some f-model) of an infinite game term is the lub of the semantics of the members of its ideal,
if the lub exists in the f-model.
\end{defi}

\begin{defi}
Let $M\typ \si$ be a closed PCF-term.\\
$\FP^\si_0 M \lsy \FP^\si_1 M \lsy \FP^\si_2 M \lsy \ldots$
is an ascending chain of finite terms with ascending grade.\\
Define $\gts(M)$ as the lub (in the order of infinite game terms)
of the ascending chain of game terms
$\gt^\si_0(\FP^\si_0 M) \lsy \gt^\si_1(\FP^\si_1 M) \lsy \gt^\si_2(\FP^\si_2 M) \lsy\ldots$.
\end{defi}

\begin{thm}[Infinite Game Term Theorem]
If $M\lsy N\typ \si$ are closed PCF-terms,
then $\gts(M)\lsy \gts(N)$ are infinite game terms with
$\sem{M} = \sem{\gts(M)}$ and $\sem{N} = \sem{\gts(N)}$ in any f-model.
\end{thm}
\proof
By proposition \ref{p:projection} it is $\sem{\FPsi M} \flub \sem{M}$,
therefore $\sem{M} = \sem{\gts(M)}$,
and likewise $\sem{N} = \sem{\gts(N)}$.
As $\gti(\FPsi M) \lsy \gti(\FPsi N)$ for all $i$,
we get $\gts(M) \lsy \gts(N)$.
\qed

\section{The syntactic order is not the image of the stable order}

\newcommand{\Dnn}{D^n_n}
\newcommand{\Dbnn}{\bar{D}^n_n}
\newcommand{\gn}{g_{n+1}}
\newcommand{\lagn}{\la \gn \ldots g_1.}
\newcommand{\pqz}{p\ldots p q\ldots q \mto \0}

\newcommand{\D}{%
\tlf{D = \la g.}{g}%
{\tz \tf{g}{\ti \ti}{\tbo}{\tz}}%
{\tz}{\tbo}%
}
\newcommand{\C}{%
\tlf{C = \la g.}{g}%
{\tbo \tf{g}{\ti \ti}{\tbo}{\tz}}%
{\tz}{\tbo}%
}
\newcommand{\B}{%
\tlf{B = \la g.}{g}%
{\tbo \tf{g}{\ti \ti}{\tbo}{\tz}}%
{\longarm{\tf{g}{\ti \ti}{\tz}{\tz}}}{\treesp{-4mm}\tbo}%
}
\newcommand{\A}{%
\tlf{A = \la g.}{g}%
{\tbo \tf{g}{\ti \ti}{\tbo}{\tz}}%
{\longarm{\tf{g}{\ti \ti}{\tbo}{\tz}}}{\treesp{-4mm}\tbo}%
}
\newcommand{\Dnni}{%
\tlf{\bar{D}^{n+1}_{n+1} ={}}{\gn}%
{\tz \tf{\gn}{\ti \ti}{\tbo}{\tz}}%
{\tc{\Dbnn}}{\tbo}%
}
\newcommand{\Cnni}{%
\tlf{\bar{C}^{n+1}_{n+1} = }{\gn}%
{\tbo \tf{\gn}{\ti \ti}{\tbo}{\tz}}%
{\tc{\Dbnn}}{\tbo}%
}
\newcommand{\Din}{%
\tlf{\bar{D}^i_{n+1} = }{\gn}%
{\tbo \tf{\gn}{\ti \ti}{\tbo}{\tz}}%
{\longarm{\tf{\gn}{\ti \ti}{\tc{\bar{D}^i_n}}{\tc{\Dbnn}}}}{\treesp{-4mm}\tbo}%
}
\newcommand{\Cin}{%
\tlf{\bar{C}^i_{n+1} = }{\gn}%
{\tbo \tf{\gn}{\ti \ti}{\tbo}{\tz}}%
{\longarm{\tf{\gn}{\ti \ti}{\tc{\bar{C}^i_n}}{\tc{\Dbnn}}}}{\treesp{-4mm}\tbo}%
}

Berry's second conjecture in its finite form says that the stable order of the order-extensional
fully abstract cpo-model of PCF (our greatest f-model) has the syntactic order as its image:\\
If $a\lst b$ for finite $a,b$ in the model, then there are normal form terms $A,B$
with $\sem{A} =a$, $\sem{B}=b$ and $A\lsy B$.\\
(The choice of the greatest f-model is not important, as all f-models coincide on their finite parts.)

In this section we will first show that Berry's second conjecture is valid in first-order types.
Then we give our simplest counter-example in finitary PCF of second-order type,
a chain of length 2.
We also give examples of chains of any finite length.

For first-order types Berry's conjecture can be strengthened to the infinite case:

\begin{thm}[Berry, Theorem 4.1.7 and  4.8.14 in \Berry] \label{t:firstimage}
Let $\si$ be a first-order type, and $b\in \Ds$ in the greatest f-model.
Then there is an infinite game term $B$  with $b=\sem{B}$.
Furthermore, for all such infinite game terms $B$ and every subset $t\sub \trace(b)$
there is an infinite game term $A\lsy B$ with $\tsem{A} = t$.
(As infinite game term, $A$ has a denotation in the greatest f-model.)
\end{thm}
\proof

Let $\si = \io\fun\io\fun \ldots \fun \io$ with $n\geq 1$ arguments.
In \cite[4.1.7]{Berry}
 Berry shows that $b\in \Ds$, as the lub of a growing sequence of finite sequential
functions, is itself sequential.
Therefore: If $b$ is not some constant function, then $b$ is strict in some $j$-th argument.
So $B$ can be recursively constructed as infinite game term (with $\caseinf$ the infinite $\case$)
in the form:
\[ B = \laxn \caseinf x_j B_1 B_2 \ldots,\]
where $B_i$ is a term with free variables $x_1,\ldots x_{j-1} x_{j+1} \ldots x_n$
for the residual function $b_i$ given by
\[ b_i x_1\ldots x_{j-1} x_{j+1}\ldots x_n = b x_1 \ldots x_{j-1} i x_{j+1} \ldots x_n .\]

In \cite[4.8.14]{Berry} Berry shows that $A$ can be constructed in the same manner $B$ was constructed,
\ie following the same choice of the variables for which the function is strict.
We can describe the construction of $A$ differently by using traces:
The tokens of the trace $\tsem{B}$ correspond exactly to the branches of $B$ that output a result,
\ie do not lead to $\Om$.
We simply choose $A\lsy B$ by setting those branches of $B$ that do not correspond to a token in
$t$ to the empty output $\Om$.
\qed

We conjecture that Berry's second conjecture is also true for second-order types
with parameters of arity at most one:
\begin{conj} \label{c:aritysecond}
Let $\si=\sinio$ with $\si_i = \io$ or $\si_i = \io\fun \io$ for all $i$.
Let $b\in \finsi$ be a finite element of grade $i$.\\
Then there is a game term $B$ of grade $i$ for $b$, $b=\sem{B}$,
such that for every subset $t\sub \trace(b)$ that is secured in the sense of definition 2
of \cite{Curien/Plotkin} there is $A\lsy B$ with $\tsem{A}=t$.
(The trace of every semantic element is secured, so Berry's second conjecture would be fulfilled for these types.)
\end{conj}
The proof of this conjecture is in preparation.
It needs a new theory of (PCF-)terms that would exceed the frame of this paper.

\subsection{Refutation of Berry's second conjecture: A chain of least length 2} \label{s:firstexample}

Our simplest counter-example to Berry's second conjecture is in finitary PCF
of second-order type $(\io\fun\io\fun\io)\fun\io$.
We consider the following game terms $A,B,C,D$:

\bigskip

\hspace{3mm}  \tree{\D}  \hspace{40mm} \tree{\B}

\hspace{3mm}  \tree{\C}  \hspace{40mm} \tree{\A}

\bigskip

\lstring{\D}

\smallskip

\lstring{\C}

\smallskip

\lstring{\B}

\smallskip

\lstring{\A}

\bigskip

For illustration (not for the proof) we give the trace semantics of these terms:

\medskip
\leftb{1}{2}{A}%
$\begin{aligned}[t]
\{\bi\bi \mto \bi, \tsp  \bo\bz \mto \bz\} \mto \bz \\
\{\bo\bi \mto \bi, \tsp  \bo\bz \mto \bz\} \mto \bz \\
\{\ph{\bo\bi\mto\bi,{}}\tsp \bo\bo \mto \bz\} \mto \bz \\
\{\ph{\bo\bi\mto\bi,{}}\tsp \bz\bo \mto \bz\} \mto \bz \\
\{\bi\bi \mto \bi, \tsp  \bz\bz \mto \bz\} \mto \bz \\
\{\bi\bo \mto \bi, \tsp  \bz\bz \mto \bz\} \mto \bz \\
\{\bo\bi \mto \bi, \tsp  \bz\bz \mto \bz\} \mto \bz 
\end{aligned}$%
\rightb{1}{3}{B \eq C}%
\rightb{1}{7}{D}%

\bigskip
We have $A\lsy B \eq C \lsy D$, 
therefore $\sem{A} \lst \sem{D}$.
We will prove that this chain of two steps of $\lsy$ cannot be replaced by one single step.

Proof of the equivalence $B\eq C$: For any argument $g$, if $C g$ converges
(\ie reduces to an integer constant), then the subterm $g\1\1$ of $C$ converges also.
(There are only two possibilities for $g$: either $\trace(g) = \{\bo\bo \mto \0\}$, or $g$
demands its second argument.) 
Therefore it is possible to safely replace the result $\0$ in $C$ by the term
$\caseum(g\1\1){\0}{\0}$, \ie to ``lift'' $g\1\1$ to the top level.

It is important to notice that this transformation cannot be performed with $D$:
Here there are more possibilities for $g$ to make $D g$ converge. It might be
that $\trace(g) = \{\0\bo \mto \0\}$, then the subterm $g\1\1$ does not converge.

The intuition of the example: 
We start with term $D$, working downwards step by step to $A$ eliminating tokens of the trace.
First the token $\{\0\bo \mto \0\}\mto \0$ is eliminated getting $C$
(and the other tokens with $g$ demanding its first argument $\0$).
Then it becomes possible to lift $g\1\1$, we get $B\eq C$.
Next we eliminate the token $\{\bo\bo\mto \0\}\mto\0$ in $B$ to get $A$.
This is done by ``forcing'' the evaluation of the second argument of $g$,
by demanding that $g$ delivers different results for different arguments.

\begin{prop}\label{p:AD}
Let $A,D$ be the game terms of grade $1$ above.
There are no game terms $A',D'$ of grade $1$ with $A'\lsy D'$ and $A'\eq A$, $D'\eq D$.
Then by the game term theorem \ref{t:gameterm} there are no PCF-terms $A',D'$ with this property.
 Since we have seen that $\sem{A} \lst \sem{D}$,
the proposition refutes Berry's second conjecture.
\end{prop}
\proof
As game terms of grade $1$, $A'$ and $D'$ should be of the form $\la g. S$,
where $S\typ \io$ is a game term possibly with the only free variable $g$.
We abbreviate $S[g:=M]$ as $S[M]$.

Let $R,P,Q \typ \io\fun\io\fun\io$ be the following terms:
\begin{align*}
R &= \la xy. \caseum{y}\,{\0}{(\caseum{x}\Om{\1})}, & \tsem{R} &= \{\1\1\mto\1,\tsp\bo\0\mto\0\}\\
P &= \la xy. \0,                          & \tsem{P} &= \{\bo\bo\mto\0\}\\
Q &= \la xy. \caseum{x}\,{\0}\Om,                 & \tsem{Q} &= \{\0\bo\mto\0\}
\end{align*}
We will prove: For any terms $S, S'$ of the form above,
\[ \text{if } S'\lsy S \text{ and } S[Q]\ret \0 \text{ and } S'[R]\ret\0, \text{ then }S'[P]\ret \0. \]
The proposition follows from this claim, as $DQ\ret\0$ and $AR\ret\0$, but not $AP\ret\0$.

The proof of the claim is by induction on the term $S$:\\
The cases $S=\Om,\0,\1$ are clear.\\
Let $S=\caseum(g S_1 S_2)S_3 S_4$ and $S'\lsy S$ with
 $S'=\caseum(g S'_1 S'_2)S'_3 S'_4$.
(The remaining case $S'=\Om$ is clear.)\\
Suppose $S[Q]\ret \0$ and $S'[R]\ret\0$. Then $S_1[Q]\ret \0$.\\
$R$ and $Q$ are compatible in the Scott model of all continuous functions,
the ``parallel or'' is an upper bound.
Expressed differently, $R$ and $Q$ are compatible in  the sense that they produce
compatible integer results for the same argument.
Therefore the semantics of $S_1[R]$ and $S_1[Q]$ must be compatible,
so it is not possible that $S_1[R]\ret \1$.\\
As $S'_1\lsy S_1$, it is also not possible that $S'_1[R]\ret \1$.\\
Therefore $(g S'_1 S'_2)[R]\ret \0$ (it must converge to get $S'[R]\ret \0$).\\
Hence $S'[R]\ret S'_3[R]\ret \0$.\\
On the other side we have $S[Q]\ret S_3[Q]\ret \0$.\\
Together we have $S_3[Q]\ret\0$ and $S'_3[R]\ret\0$,
and by the induction hypothesis for $S_3$ follows: $S'_3[P]\ret\0$.\\
Therefore $S'[P]\ret S'_3[P]\ret \0$.
\qed

\begin{rem}
As we base our proof on game terms, we gave a special induction
hypothesis for the combination of $\caseum$ and $g$.
The proof for general normal form terms is more complicated as it must work with $\pif$
and $g$ separately and use a more general induction hypothesis,
\ie one proves by induction on $S$:
\[ \text{If } S'\lsy S,\text{ then }\sem{S[Q]}=\sem{S'[R]}=\sem{S'[P]}\text{ or }\sem{S[Q]}=\bo
\text{ or } \sem{S'[R]}=\bo \]
This has on the surface the \emph{form} of the Sieber sequentiality logical relation
$S^3_{\{1,2\}\{1,2,3\}}$, see \cite{Sieber}.
(It is $(d_1,d_2,d_3)\in 
S^3_{\{1,2\}\{1,2,3\}}$ iff $d_1=d_2=d_3$ or $d_1=\bo$ or $d_2=\bo$.)
This form on the surface is responsible for the fact that the induction hypothesis goes up through
the case $S= \pif S_1 \pthen S_2 \pelse S_3$.
But for the proof of the case $S=gS_1S_2$
the specific semantics of $R,P,Q$ and the fact $S'\lsy S$ are needed.

So a sequentiality relation alone is not sufficient to prove this counter-example:
a logical relation is a \emph{semantic} means to prove the undefinability of a function.
But here we must prove the undefinability of $S'\lsy S$
for two functions $\sem{A}\lst \sem{D}$,
where both functions separately are definable.
At first sight this necessitates a \emph{syntactic} proof.
But we could ask the question:
Are there \emph{semantic} means to prove this?
Are there necessary semantic conditions for the syntactic order that are
stronger than the condition of stable order?
See also the remark in the last section ``Outlook''.
\end{rem}

\subsection{Chains of any length}

We have seen an example of a chain of two $\lsy$-steps.
Generally:

\begin{defi} \label{d:chain}
Let $a\lst b$ be finite elements in an f-model.\\
A \defin{chain of length} $n\geq 1$ \defin{between $a$ and $b$} is a pair of sequences
of terms $(C_i), (D_i)$ with $1\leq i \leq n$ and $a=\sem{C_1}$, $b=\sem{D_n}$ and
$C_i \lsy D_i$, $D_i\eq C_{i+1}$.\\
If $a=b$, then we say there is a chain of length $0$ between $a$ and $b$.\\
A chain is  \defin{of least length} $n$ if there is no shorter chain.
\end{defi}

By the game term theorem, if there is a chain of PCF-terms, then there is an equivalent chain of
game terms.

Now we construct examples of chains of least length $n+1$ for any finite $n\geq 0$,
by a sequential composition of $n$ copies of our first example, each copy for a different
argument $g_i$.
For every $n\geq 0$ let $\si_n$ be the type $(\io\fun\io\fun\io)\fun \ldots \fun (\io\fun\io\fun\io)\fun\io$
with $n$ parameters.
For every $n$ we define two sequences of game terms $C^i_n, D^i_n \typ \si_n$ with $0\leq i \leq n$.\\
First we define by induction on $n$  the versions $\bar{C}^i_n, \bar{D}^i_n$ without $\la$-binder:

\no
\tc{\bar{D}^0_0 = 0}  \hfill  \Dnni  \hfill \Din \quad for $i\leq n$  

\no
\tc{\bar{C}^0_0 = \Om}  \hfill \Cnni \hfill  \Cin \quad for $i\leq n$  

\vspace{5mm}

We define $C^i_n= \la g_n\ldots g_1. \bar{C}^i_n$
and $D^i_n= \la g_n\ldots g_1. \bar{D}^i_n$.\\
For all $n\geq 0$, $0\leq i\leq n$: $C^i_n\lsy D^i_n$.
The proof is an easy induction on $n$.\\
For all $n\geq 1$, $i<n$: $D^i_n \eq C^{i+1}_n$.
Proof by induction on $n$:\\
For $n=1$, $i=0$ we have that $D^0_1$ is the term $B$, and $C^1_1$ the term $C$ of our former example,
both only with $g$ replaced by $g_1$.\\
For $n:=n+1$:\\
For $i=n$ we have $D^n_{n+1} \eq C^{n+1}_{n+1}$ by the same argument as in our former example
for $B\eq C$.\\
For $i<n$ we get $D^i_{n+1} \eq C^{i+1}_{n+1}$ by the induction hypothesis.

All together for any $n\geq 0$ we get a chain of length $n+1$ between $\sem{C^0_n}$
and $\sem{\Dnn}$:
\[ C^0_n \lsy D^0_n \eq C^1_n \lsy D^1_n \ldots D^{n-1}_n \eq C^n_n \lsy \Dnn. \]
We want to prove that this chain has the least length.

\bigskip

First the intuition of the example:
We use the terms $R,P,Q$ of the proof of proposition \ref{p:AD} and name their traces:
\[ r=\tsem{R}=\{\1\1\mto\1, \bo\0\mto\0\},\quad p=\tsem{P}=\{\bo\bo\mto\0\},\quad q=\tsem{Q}=\{\0\bo\mto\0\} \]
The trace of $\Dnn$ contains all tokens $\pqz$, with $j$ arguments $p$, $0\leq j \leq n$.
These tokens are in the upper branch of $\Dnn$.
We work down from $\Dnn$ eliminating all these tokens in $n+1$ steps.

In the $j$-th step ($0\leq j \leq n$) the token $\pqz$, with $j$ arguments $p$,
is eliminated in $D^{n-j}_n$.
(In $D^{n-j}_n$ all the tokens of this form with less arguments $p$ have already been eliminated.)
If $j < n$ we proceed as follows:
Following the upper branches in $D^{n-j}_n$ we come to an occurrence of the variable $g_{n-j}$.
It is the root of a subterm $\bar{D}^{n-j}_{n-j}$, its upper arm is $\bar{D}^{n-j-1}_{n-j-1}$.
The elimination is by setting the first argument of this $g_{n-j}$ to $\Om$, getting $C^{n-j}_n$.
Only then it is possible to lift the lower $g_{n-j}\1\1$ to the top level, getting $D^{n-j-1}_n$.
There the new $g_{n-j}\1\1$ at the top level gets two arms which are copies of $\bar{D}^{n-j-1}_{n-j-1}$.
The lower arm (of these two) stays the same in the following transformations
(it contains the token $p\ldots p r q\ldots q\mto\0$ with $j$ arguments $p$).
The upper arm undergoes further eliminations of tokens $\pqz$.
These further eliminations are only possible after the separation of the two arms.

Finally in the $n$-th step the $\0$ which stands at the end of the upper branches of $D^0_n$
is set to $\Om$ getting $C^0_n$, eliminating the token $p\ldots p\mto \0$.

\begin{prop}
Let $n\geq 0$ and $C^i_n, D^i_n$ be the terms defined above.
Then the chain
\[ C^0_n \lsy D^0_n \eq C^1_n \lsy D^1_n \ldots D^{n-1}_n \eq C^n_n \lsy \Dnn \]
between $\sem{C^0_n}$ and $\sem{\Dnn}$
has the least length $n+1$.
\end{prop}
\proof
We assume $n\geq 1$ and
suppose any chain between $C^0_n$ and $\Dnn$ and look at an intermediate $\lsy$-step of this chain,
\ie we have the situation
\[ C^0_n \lst M \lsy N \lst \Dnn.\]
We assume that some token of the form $\pqz$ is eliminated in this step.
Let $t$ be such token with the minimal number $j$ of arguments $p$,
and assume $j<n$.\\
Then we have
\[ NP\ldots P Q\ldots Q \ret \0,\text{ and }M P\ldots PRQ\ldots Q\ret \0, \]
because $C^0_n \lst M$ (both with $j$ arguments $P$).\\
We can abstract the $(j+1)$st argument in these terms and build the terms
\[ N'=\la g.NP\ldots PgQ\ldots Q \text{ and } M'=\la g. MP\ldots PgQ\ldots Q .\]
It is $M'\lsy N'$.
We can transform $M',N'$ to game terms and apply the argument in the proof of proposition \ref{p:AD}
to deduce: 
$M'P\ret \0$.\\
So $MP\ldots PPQ\ldots Q \ret \0$ (with $j+1$ arguments $P$).\\
As $Q \lop P$, we also have $MP\ldots PQ\ldots Q\ret \0$ for all $k\geq j+1$ arguments $P$.\\
All these arguments of $M$ are minimal \wrt the stable order,
because they are also minimal for $\Dnn$ and it is $M\lst \Dnn$.\\
Therefore every token $\pqz$ with $k\geq j+1$ arguments $p$ is in $M$.\\
This shows that from the tokens of the form $\pqz$ only the token $t$ is eliminated in the
step $M\lsy N$.
(For $j=n$ this is trivially the case.)
As there are $n+1$ of these tokens to be eliminated, the chain must have at least $n+1$ steps.
\qed

Our example of a chain of least length $n+1$ has $n$ functional parameters $g_i$
of arity $2$ and is of grade $1$.
We could transform it into an ``equivalent'' example with only one functional parameter $g$
of arity $3$ and terms of grade $n$,
by coding $g_i MN$ as $giMN$.

Our results suggest an improvement of Berry's second conjecture:
\begin{conj}[Chain Conjecture] \label{c:chain} 
If $a\lst b$ are finite elements in an f-model,
then there is a chain between $a$ and $b$.\\
We will refute also this conjecture in section \ref{s:chain}.
\end{conj}

\section{The stable order is not bounded complete: no bidomain}

\newcommand{\Aa}{\tlf{\bar{A}=}{g}{\ti \ttwo}{\tbo}{\longarm{\tf{g}{\tz\tbo}{\tz}{\tbo}}}}
\newcommand{\Ba}{\tlf{\bar{B}=}{g}{\ti \ti}{\tbo}{\longarm{\tf{g}{\tbo\tz}{\tz}{\tbo}}}}
\newcommand{\Ca}{\tlf{C=\la g.}{g}{\tc{\bar{A}} \tc{\bar{B}}}{\tz}{\tbo}}

\newcommand{\Ab}{\tlf{\bar{A}=}{g}{\ti \tz \ti}{\tbo}{\longarm{\tf{g}{\tz\tbo\tbo}{\tz}{\tbo}}}}
\newcommand{\Bb}{\tlf{\bar{B}=}{g}{\ti \ti \ti}{\tbo}{\longarm{\tf{g}{\tbo\tbo\tz}{\tz}{\tbo}}}}
\newcommand{\Cb}{\tlf{C=\la g.}{g}{\tc{\bar{A}} \tbo \tc{\bar{B}}}{\tz}{\tbo}}

\newcommand{\treespa}{\treesp{-11mm}}

\newcommand{\Ac}{\tlf{\bar{A}=}{f}{\ti \ttwo}{\tbo\treespa}{\longarm{\tf{f}{\tz\tbo}{%
\longarm%
{\tf{g}{\ti \ttwo}{\tbo}{\longarm{\tf{g}{\tz\tbo}{\tz}{\tbo}}}}
}{\treespa\tbo}}}}

\newcommand{\Bc}{\tlf{\bar{B}=}{f}{\ti \ti}{\tbo\treespa}{\longarm{\tf{f}{\tbo\tz}{%
\longarm%
{\tf{g}{\ti \ti}{\tbo}{\longarm{\tf{g}{\tbo\tz}{\tz}{\tbo}}}}
}{\treespa\tbo}}}}

\newcommand{\Cc}{\tlf{C=\la fg.}{f}{\tc{\bar{A}}  \tc{\bar{B}}}{\tz}{\tbo}}
\newcommand{\Dc}{\tlf{D=\la fg.}{g}{\tc{\bar{A}}  \tc{\bar{B}}}{\tz}{\tbo}}
\newcommand{\Ec}{\tlf{E=\la fg.}{f}{\tc{\bar{A}}  \tc{\bar{B}}}{%
\longarm%
{\tf{g}{\tc{\bar{A}} \tc{\bar{B}}}{\tz}{\tbo}}
}{\treespa\tbo}}

G\'erard Berry showed that the fully abstract order-extensional cpo-model of PCF (our greatest f-model)
together with the stable order forms a bicpo,
and conjectured that it is also a bidomain (Berry's first conjecture).
Here we repeat the definitions of both structures.
We prove the conjecture for first-order types.
Then we refute the general conjecture.
Our first example is the stable lub of two finite elements for which the distributive law is not valid.
Our second example consists of two finite elements with stable upper bound but without stable lub.
Both examples are in PCF of second-order type of grade $2$.

\begin{defi}[Berry: 4.7.2 in \Berry]
A \defin{bicpo} is a structure $(D,\lex,\lst,\bo)$ such that:
\begin{enumerate}[(1)]
\item The structure $(D,\lex,\bo)$ is a cpo with least element $\bo$ and with
a continuous glb-function $\glbex$.
\item The structure $(D,\lst,\bo)$ is a cpo with least element $\bo$ such that
$a\lst b \ifthen a\lex b$ and
for all $\lst$-directed sets $S$ the two lubs are equal: $\Lubst S = \Lubex S$.
\item The function $\glbex$ is $\lst$-monotonic.
(With (1) and (2) it follows that it is $\lst$-continuous.)
\item For all $\lst$-directed sets $S$ and $S'$:
If for all $a\in S$, $a'\in S'$ there are $b\in S$, $b'\in S'$ with $a\lex b$, $a'\lex b'$, $b\lst b'$,
then $\Lubex S \lst \Lubex S'$.
\end{enumerate}
\end{defi}

In a bicpo: For all $a\cost b$, $a\glbex b$ is also the glb \wrt $\lst$.

\begin{thm}[Berry: 4.8.10 in \Berry]
The domains $(\Ds,\lex,\lst,\bo)$
of the fully abstract order-extensional cpo-model of PCF are bicpos.
\qed
\end{thm}

\begin{defi}[Berry: 4.4.10 in \Berry]
A cpo $(D,\lst,\bo)$ is \defin{distributive} if
\begin{enumerate}[(1)]
\item it is bounded complete\\
(This means that for $a\cost b$ there is a lub $a\lubst b$.
And this entails with completeness that there is also a glb $a\glbst b$ for all $a,b$,
even for $\lst$-incompatible ones.)\\
and
\item for all $a,b,c \in D$ with $b\cost c$:  $a\glbst(b\lubst c) = (a\glbst b)\lubst(a\glbst c)$.
\end{enumerate}
\end{defi}

\begin{defi}[Berry: 4.7.9 in \Berry]
A bicpo $(D,\lex,\lst,\bo)$
is \defin{distributive} if $(D,\lst,\bo)$ is distributive
and for all $a\cost b$: $a\lubst b$ is also the lub \wrt $\lex$.
\end{defi}

(Please note that in a distributive bicpo only for $a\cost b$ it must be $a\glbst b = a\glbex b$.)

\begin{defi}[Berry: 4.7.12 in \Berry]
A distributive bicpo $(D,\lex,\lst,\bo)$ is a \defin{bidomain}
if there is a $\lst$-growing sequence $(\psi_i)_{i\geq 1}$
of finite projections \wrt $\lst$ and with lub $\Lubst \psi_i = \id$.\\
(This means: $\psi_i\typ D\fun D$ is continuous \wrt $\lex$ and $\lst$,
$\psi_i\lst\id$, $\psi_i\comp \psi_i = \psi_i$,
$\psi_i\lst \psi_{i+1}$, $\psi_i(D)$ finite, $\Lubst \psi_i = \id$.)
\end{defi}

In this definition the sequence $(\psi_i)$ is also a $\lex$-growing sequence of finite
projections \wrt $\lex$ and with lub $\id$.
Together with the the glb-function $\glbex$ it follows that $(D,\lex,\bo)$ is a Scott domain,
a bounded complete $\omega$-algebraic cpo.

As we have explained in proposition \ref{p:projection} and \ref{p:propI},
the conditions for $(\psi_i)$ in the definition of bidomain are fulfilled
for the fully abstract order-extensional cpo-model (and furthermore for all f-models)
by the projections $\fpsi$.
In fact the $(\Ds,\lst)$ are stable $\omega$-bifinite domains for the cpo-model,
in the sense of definition 12.4.3 of \cite{Amadio/Curien}.

To be precise, the condition of distributivity of the stable order was not conjectured by Berry
in his thesis; there he remained agnostic.
But in the state-of-the-art paper \cite{Berry/Curien} we can read:
``Unfortunately we are not able to show that the domains of the fully abstract model are bidomains,
although we definitely believe it;
the problem is to show that the $\lst_{cm}$-lubs are taken pointwise.''

First we clarify the situation for first-order types:

\begin{thm}
Let $\si$ be a first-order type and $(\Ds,\lex,\lst,\bo)$ be the corresponding domain of any f-model.\\
The finite elements of $\Ds$ fulfill distributivity \wrt $\lst$ in $\Ds$ in the following sense:\\
For $a,b\in \fins$ the glb in $\Ds$ exists and is given by $\trace(a\glbst b)=\trace(a)\cap\trace(b)$.\\
For $a,b\in \fins$ with $a\cost b$ the lub in $\Ds$ exists and is given by
 $\trace(a\lubst b)=\trace(a)\cup\trace(b)$.
It is taken pointwise and it is also the lub \wrt $\lex$.\\
Then the distributive law is fulfilled by set theory on traces.

If $\Ds$ contains a denotation for every infinite game term of type $\si$
(this is the case for the game model and every greater f-model),
then $\Ds$ is the domain of the greatest f-model.
In this case all elements $a,b\in \Ds$ fulfill distributivity in the sense above.
Therefore $\Ds$ is a bidomain in this case.
\end{thm}
\proof
Let $a,b\in \fins$.\\
We can apply theorem \ref{t:firstimage} and get a game term $A$ with $a=\sem{A}$,
and a game term $C\lsy A$ with $\tsem{C}= \trace(a)\cap\trace(b)$.
Define $a\glbst b = \sem{C}$; it is finite and therefore in $\Ds$.

Now let $a\cost b$, \ie there is some $d$ with $a\lst d$ and $b\lst d$.
By theorem \ref{t:firstimage} there are an infinite game term $D$ with $d=\sem{D}$,
and finite game terms $A,B$ with $a=\sem{A}$, $b=\sem{B}$, $A\lsy D$, $B\lsy D$.
Take the syntactical lub $E$ of $A$ and $B$.
It is $\tsem{E} = \tsem{A} \cup \tsem{B}$,
because in first-order game terms branches correspond to tokens.
Define $a\lubst b = \sem{E}$; it is finite and therefore in $\Ds$.
This lub is pointwise on the uncurried argument and therefore also the lub \wrt $\lex$.

If $\Ds$ contains a denotation for every infinite game term of type $\si$,
then by theorem \ref{t:firstimage} $\Ds$ is exactly the domain of the greatest f-model.
The construction of $a\glbst b$ and $a\lubst b$ for any $a,b\in \Ds$ is as above, only 
with infinite game terms.
\qed

\begin{conj} \label{c:arityfirst}
For all types of the form $\si=\sinio$, with $\si_i=\io$ or $\si_i=\io\fun\io$,
Berry's first conjecture is valid, \ie $\Ds$ is a bidomain in the greatest f-model.
\end{conj}
The proof of this conjecture is in preparation.
It relies on the conjecture \ref{c:aritysecond}.

Now we prove some properties of stable upper bounds (sub) in f-models.
(These are properties that are also valid in stable bifinite domains,
see lemma 12.4.7 in \cite{Amadio/Curien}.)

\begin{thm} \label{t:sub}
Let $\Ds$ be a domain of an f-model,
$\si=\sinio$, $n\geq 0$.
Let $X$ be a finite set of finite elements of $\Ds$ that has a stable upper bound (sub) in $\Ds$.
Let $m$ be the maximal grade of the elements of $X$.
For every sub $x$ of $X$ there is a unique minimal (\wrt $\lst$) sub $y$ of $X$ with $y\lst x$.
Every minimal sub of $X$ is finite of grade $m$;
they are pairwise $\lst$-incompatible.
The extensional lub $\Lubex X$ is one of those.
\end{thm}
\proof
Let $x$ be a sub of $X$.
Then the projection $\fp^\si_m x$ is also a sub of $X$.
Let $Z$ be the set of all subs $z$ of $X$ with $z\lst \fp^\si_m x$;
it is a non-empty finite set of finite elements.
Then $y=\Glbex Z$ is the desired unique minimal sub of $X$ with $y\lst x$.

Let $a,b$ be two minimal subs of $X$ that are $\lst$-compatible.
Then $a\glbex b$ is also a sub of $X$, therefore $a=b$.

Let $g=\Lubex X$ and $h$ some sub of $X$.
We have to show that $f\lst g$ for every $f\in X$.\\
This is clear for $n=0$, in the type $\io$.\\
Now let $n>0$ and $\vecx$, $\vecy$ be two vectors of arguments of type $\si_1\times \ldots \times \si_n$
with $\vecx\lst \vecy$.\\
We have to show that $f\vecx = f\vecy \glbex g\vecx$.\\
It is $f\vecx = f\vecy \glbex h\vecx \gex f\vecy \glbex g\vecx$.
And $f\vecx \lex f\vecy \glbex g\vecx$ is clear.\\
This shows that $g$ is a sub of $X$; of course it is also minimal \wrt $\lst$.
\qed

\subsection{A stable lub without distributivity} \label{s:distrib}

Our first counter-example to Berry's first conjecture is of type $(\io\fun\io\fun\io)\fun\io$
and of grade $2$.
We consider the following game terms $A,B,C$,
where we use a $\caseum$ for a $\caset$ with the third arm $\Om$:

\Aa \hfill \Ba \hfill \Ca

\no
$A =\la g.\bar{A}$ \hspace{47mm}  
$B =\la g.\bar{B}$  

\medskip
\no
Here are the traces of these terms:

\leftb{1}{2}{A}%
\hspace{-7mm}\leftb{3}{4}{B}%
$\begin{aligned}[t]
\{\bz\bo\mto\bz,\tsp\bi\bt\mto\bi\}\mto\bz\\
\{\bz\bo\mto\bz,\tsp\bi\bo\mto\bi\}\mto\bz\\
\{\bo\bz\mto\bz,\tsp\bi\bi\mto\bi\}\mto\bz\\
\{\bo\bz\mto\bz,\tsp\bo\bi\mto\bi\}\mto\bz\\
\{\bo\bo\mto\bz\ph{{},\tsp\bo\bi\mto\bi}\}\mto\bz
\end{aligned}$%
\rightb{1}{5}{C}

\medskip

It is $A\lst C$ and $B\lst C$.
We will show that $C$ is the stable lub of $A$ and $B$.

The intuition of the example:
$A$ and $B$ do not contain the token $\{\bo\bo\mto\0\}\mto\0$,
because their two occurrencies of $g$ are forced to evaluate their first \resp second
argument, to get different results for different arguments.
(This is the same trick that was used in the preceding section.)
$C$ adds to the tokens of $A$ and $B$ just the token $\{\bo\bo\mto\bz\}\mto\0$,
to separate $\bar{A}$ and $\bar{B}$.
(Note that a $g$ for which $Cg$ converges cannot demand both its arguments $\0\0$.)
Therefore this lub does not fulfill distributivity.
In $C$ it is not possible to lift a differing term $gMN$ to the top level
that would eliminate that token,
because the five occurrences of $g$ in $C$ cannot be ``unified''
to a common term that would always converge.

\begin{prop}\label{p:ABC}
Let $A,B,C$ be the game terms above.
$\sem{C}$ is the stable lub of $a=\sem{A}$ and $b=\sem{B}$.
Let $d$ be the finite element with the trace $\{\{\bo\bo\mto\0\}\mto\0\}$.
Then $d\glbst(a\lubst b) \neq (d\glbst a)\lubst(d\glbst b)$.
This refutes Berry's first conjecture.
\end{prop}
\proof
By the game term theorem \ref{t:gameterm} and the preceding theorem \ref{t:sub},
every minimal sub of $A$ and $B$ can be represented by a game term of grade $2$.
Such a  game term is of the form $\la g.S$, where $S\typ \io$ is a game term
possibly with the only free variable $g$.
We abbreviate $S[g:=M]$ as $S[M]$.

We use the following terms as arguments:
\begin{align*}
Q &= \la xy.\caseum x\0(\caset y \bo\bo\1) & \tsem{Q} &= \{\0\bo\mto\0,\tsp \1\2\mto\1\} \\
R &= \la xy.\caseum y\0(\caseum x \bo\1) & \tsem{R} &= \{\bo\0\mto\0,\tsp \1\1\mto\1\} \\
P &= \la xy. \0   &  \tsem{P} &= \{\bo\bo\mto\0\}
\end{align*}
$Q$ and $R$ are compatible in the sense that they produce compatible results for the same argument.
We will prove that 
for any term $S$ of the form above:
\[ \text{If } S[Q]\ret \0 \text{ and } S[R]\ret\0, \text{ then } S[P]\ret\0.\]
The proof is by induction on the term $S$:
The cases $S=\Om,\0,\1,\2$ are clear.\\
Let $S=\caset(gS_1S_2)S_3S_4S_5$.\\
For $S[Q]\ret\0$ it must be $S_1[Q]\ret\0$ or $S_2[Q]\ret\2$.
\begin{enumerate}[(1)]
\item case $S_1[Q]\ret\0$:\\
For $S[R]\ret\0$ it must be $S_2[R]\ret\0$ or $S_1[R]\ret\1$.
\begin{enumerate}[(\theenumi.1)]
\item case $S_2[R]\ret\0$:\\
We have $S[Q]\ret S_3[Q]\ret\0$ and $S[R]\ret S_3[R]\ret\0$.\\
By the induction hypothesis for $S_3$ we get $S_3[P]\ret\0$, therefore $S[P]\ret\0$.
\item case $S_1[R]\ret\1$:\\
This is not possible, as $Q$ and $R$ are compatible in the sense above.
\end{enumerate}
\item case $S_2[Q]\ret\2$:\\
For $S[R]\ret\0$ it must be $S_2[R]\ret\0$ or $S_2[R]\ret\1$.\\
Both cases are not possible, as $Q$ and $R$ are compatible in the sense above.
\end{enumerate}
So we have shown that for every $\lex$-upper bound $D$ of grade $2$ of $A$ and $B$ it must be
$DP\ret\0$.
For a $\lst$-upper bound it cannot be $D\Om\ret\0$.
Therefore $P$ is a $\lst$-minimal argument to fulfill $DP\ret\0$.
This means: Any minimal stable upper bound of $A$ and $B$ must contain the token $\{\bo\bo\mto\0\}\mto\0$.
So $C$ is the stable lub of $A$ and $B$.
(It is also the $\lex$-lub.)
\qed 

\begin{rem}[alternative proof with Sieber sequentiality relation]
Because we work in the proof above on game terms,
the induction hypothesis is simpler and the proof shorter than a proof by induction on general terms.
A short purely semantic proof for general terms is possible with a Sieber 
sequentiality logical relation \cite{Sieber}.

We can show that there is no definable function that fulfills the value table
$\sem{Q}\mto\0$, $\sem{R}\mto\0$, $\sem{P}\mto n$ for $n\neq \0$.
We use the sequentiality relation $rel = S^3_{\{1,2\}\{1,2,3\}}$.\\
For $d_1,d_2,d_3\typ \io$ it is $(d_1,d_2,d_3) \in rel$ iff $d_1=\bo$ or $d_2=\bo$ or $d_1=d_2=d_3$.\\
First, the output column $(\0,\0,n)$ of the value table is not in this relation.\\
Then we have to show that $(\sem{Q},\sem{R},\sem{P}) \in rel$ (on the type $\io\fun\io\fun\io$).\\
Suppose we have
\begin{align*}
\sem{Q} a_1 b_1 &= c_1,\\
\sem{R} a_2 b_2 &= c_2,\\
\sem{P} a_3 b_3 &= c_3
\end{align*}
and suppose $(c_1,c_2,c_3)\not\in rel$.
We have to show that $(a_1,a_2,a_3)\not\in rel$ or
$(b_1,b_2,b_3)\not\in rel$.

\no It must be $c_3=0$.\\
It cannot be $c_1=\bo$, so it must be $c_1=0$ or $c_1=1$:\\
If $c_1=0$, then it cannot be $c_2=0$,
so it must be $c_2=1$,
then $a_1=0$, $a_2=1$,
therefore $(a_1,a_2,a_3)\not\in rel$, end of proof for $c_1=0$.\\
If $c_1=1$, then it is $c_2=0$ or $c_2=1$:\\
If $c_2=0$, then $b_1=2$, $b_2=0$, therefore $(b_1,b_2,b_3)\not\in rel$.\\
If $c_2=1$, then $b_1=2$, $b_2=1$, therefore $(b_1,b_2,b_3)\not\in rel$.

It is no surprise that we have to perform a case analysis
of similar complexity as in the proof above.
But it is interesting that the whole proof of this remark can be done mechanically by the
computer program written by Allen Stoughton \cite{Stoughton:mechanizing}.
For a general system of ground constants, this program takes a value table
of a second-order function and returns either a term defining such a function
or a logical relation proving its undefinability.
\end{rem}

Our counter-example is of grade $2$ with $g$ of arity $2$.
There is an ``equivalent'' example of grade $1$ with $g$ of arity $3$:

\Ab \hfill \Bb \hfill \Cb

\no
$A =\la g.\bar{A}$ \hspace{48mm}  
$B =\la g.\bar{B}$  

\begin{conj}
In $\fin^{(\io\fun\io\fun\io)\fun\io}_1$, the finite elements of grade $1$ of the type
$(\io\fun\io\fun\io)\fun\io$, Berry's first conjecture is valid; this subdomain is a bidomain.
(This is a finite combinatorial problem and could be solved by a computer program.)
\end{conj}

\subsection{Two elements without stable lub}

Now to our counter-example to bounded completeness of the stable order.
It is of type $(\io\fun\io\fun\io)\fun(\io\fun\io\fun\io)\fun\io$ and of grade $2$.
It employs the trick of our last example twice to two functional parameters.
Consider the following game terms $A,B,C,D,E$, where we use a $\caseum$ for a $\caset$ with the
third arm $\Om$.

\Ac  

\no
$A=\la fg.\bar{A}$  

\Bc

\no
$B=\la fg.\bar{B}$  

\hspace{5mm} \Cc  \hfill \Dc  \hfill  \Ec

\medskip
\no
The traces of the terms are:
\begin{align*}
\tsem{A} &= \{\bz \bo\mto\bz, \bi\bt\mto\bi\} \mto \{\bz\bo\mto\bz, \bi\bt\mto\bi\}\mto\bz \\
 &\ph{{}= \{\bz \bo\mto\bz, \bi{}}\bo\ph{\mto\bi\} \mto \{\bz\bo\mto\bz, \bi}\bo \\
\tsem{B} &= \{\bo \bz\mto\bz, \bi\bi\mto\bi\} \mto \{\bo\bz\mto\bz, \bi\bi\mto\bi\}\mto\bz \\
         &\ph{{}= \{\bo \bz\mto\bz, {}}\bo\ph{\bi\mto\bi\} \mto \{\bo\bz\mto\bz, {}}\bo \\
\tsem{C} &= \tsem{A} \cup \tsem{B} \cup \{\{\bo\bo\mto\bz\}\mto\bo\ph{\{\bo\mto\bz\}}\mto\bz\} \\
\tsem{D} &= \tsem{A} \cup \tsem{B} \cup \{\bo\ph{\{\bo\mto\bz\}}\mto\{\bo\bo\mto\bz\}\mto\bz\} \\
\tsem{E} &= \tsem{A} \cup \tsem{B} \cup \{\{\bo\bo\mto\bz\}\mto\{\bo\bo\mto\bz\}\mto\bz\} 
\end{align*}
The token of $\tsem{A}$ entails three more tokens:
(1) with the first indicated $\2$ replaced by $\bo$,
(2) with the second indicated $\2$ replaced by $\bo$,
(3) with both replaced by $\bo$.
Likewise for the token of $\tsem{B}$.
(These entailments are due to securedness, see the definition 2 of \cite{Curien/Plotkin}.)

$C,D,E$ are three stable upper bounds of $A$ and $B$;
we will show that they are just the minimal stable upper bounds.
$E$ is the $\lex$-lub of $A$ and $B$.

The intuition of the example:
In an upper bound of $A$ and $B$, both have to be separated by some function call at the top level;
because $A$ and $B$ cannot be ``unified''.
There are three ways to choose the separator: $f$ or $g$ or (both $f$ and $g$),
realized by $C,D,E$ \resp

\begin{prop}
Let $A,B,C,D,E$ be the game terms above.
$\sem{C},\sem{D},\sem{E}$ are the minimal stable upper bounds of $\sem{A}$ and $\sem{B}$.
So $\sem{A}$ and $\sem{B}$ have no stable lub.
(This again refutes Berry's first conjecture.)
\end{prop}
\proof
By theorem \ref{t:sub}, every minimal sub of $A$ and $B$ is of grade $2$.
By the game term theorem, we restrict to game terms of grade $2$.
These game terms must have the form $\la f g.S$.
We use the terms $Q,R,P$ of the proof of proposition \ref{p:ABC}.
Our claim is:
For every term $S$ of the form above,
\[ \text{if } S[f:=Q,g:=Q]\ret\0 \text{ and } S[f:=R,g:=R]\ret\0, \text{ then } S[f:=P,g:=P]\ret\0. \]
The proof of the claim is by induction on the term $S$ and follows exactly the proof of proposition 
\ref{p:ABC}.
There is only one additional case $S=\caset(f S_1S_2)S_3S_4S_5$ of the same scheme.

So we have shown that for every $\lex$-upper bound $F$ of grade $2$ of $A$ and $B$ it must be $FPP\ret\0$.
For a $\lst$-upper bound it cannot be $F\Om\Om\ret \0$.
Hence the minimal arguments to fulfill $FPP\ret\0$ must be $(P,P)$, $(P,\Om)$ or $(\Om,P)$.
This is fulfilled by $E,C,D$ respectively.
\qed

\section{Refutation and improvement of the chain conjecture} \label{s:chain}

\newcommand{\trsp}{\treesp{-6mm}}
\newcommand{\Aaa}{\tf{g}{\ti \ttwo}{\tbo\trsp}{{\tf{g}{\tz\tbo}{\ti}{\tbo}}}}
\newcommand{\Baa}{\tf{g}{\ti \ti}{\tbo\trsp}{{\tf{g}{\tbo\tz}{\ti}{\tbo}}}}
\newcommand{\Fl}{\tf{f}{\ti \tbo}{\tbo\trsp}{{\tf{f}{\tz\tz}{\tz}{\tbo}}}}
\newcommand{\Fr}{\tf{f}{\tbo \ti}{\tbo\trsp}{{\tf{f}{\tz\tz}{\tz}{\tbo}}}}
\newcommand{\Ad}{\tlf{A=\la fg.}{f}{\longarm{\Aaa} \longarm{\Baa}}{\tbo}{\longarm{\tf{f}{\tz\tz}{\tz}{\tbo}}}}
\newcommand{\Bd}{\tlf{B=\la fg.}{g}{\longarm{\tf{g}{\ti\ttwo}{\tbo\trsp}{{\Fl}}}%
\longarm{\tf{g}{\ti\ti}{\tbo\trsp}{{\Fr}}}} \tz \tbo}

The chain conjecture \ref{c:chain} said that for finite elements $a\lst b$ there is
a chain between $a$ and $b$, see the definition \ref{d:chain} of chain.
We give here a counter-example in the type $(\io\fun\io\fun\io)\fun(\io\fun\io\fun\io)\fun  \io$
of grade $2$.
Consider the following game terms $A,B$:

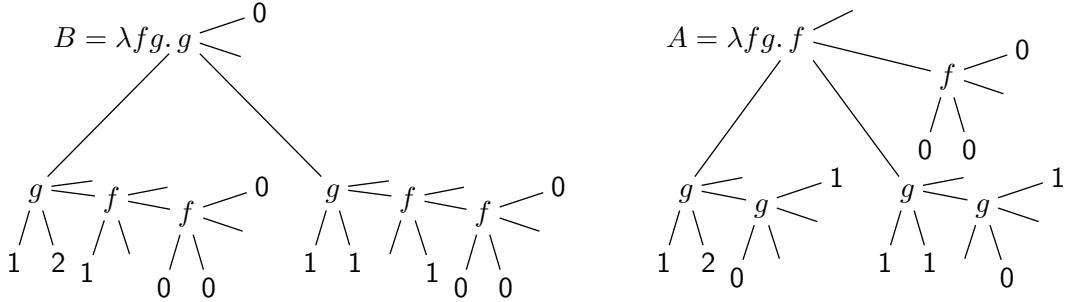
\Bd
\hspace{50mm}
\Ad

\bigskip
\no
Here are the traces of these terms:

\medskip

\leftb{1}{4}{A}%
$\begin{aligned}[t]
&\{\bz\bz\mto\bz,\tsp\bi\bo\mto\bi\}\mto\{\bz\bo\mto\bz,\tsp\bi\bt\mto\bi\}\mto\bz\\
&\ph{\{\bz}\bo\ph{\mto\bz,\tsp\bi\bo\mto\bi\}\mto\{\bz\bo\mto\bz,\tsp\bi}\bo\\
&\{\bz\bz\mto\bz,\tsp\bo\bi\mto\bi\}\mto\{\bo\bz\mto\bz,\tsp\bi\bi\mto\bi\}\mto\bz\\
&\ph{\{}\bo\ph{\bz\mto\bz,\tsp\bo\bi\mto\bi\}\mto\{\bo\bz\mto\bz,\tsp{}}\bo\\
&\ph{\{}\bo\ph{\bz\mto\bz,\tsp\bo\bi\mto\bi\}}\mto\{\bo\bo\mto\bz\ph{,\tsp\bi\bi\mto\bi}\}\mto\bz
\end{aligned}$%
\rightb{1}{5}{B}

\medskip
The first token entails three more tokens: (1) with the indicated $\0$ replaced by $\bo$,
(2) with the indicated $\2$ replaced by $\bo$,
(3) with both replaced by $\bo$.
Likewise for the second token.

It is $A\lst B$.
$B$ contains just one more token $t$ than $A$.
Assume that there is a chain between $\sem{A}$ and $\sem{B}$.
Then $t$ is eliminated in a definite step $A'\lsy B'$ of the chain, 
with $A\eq A'$ and $B\eq B'$.
We will show that such $A'\lsy B'$ do not exist.

The intuition of the example:
It is derived from the example of subsection \ref{s:distrib}.
$A$ and $B$ are like the term $C$ of that example.
For $B$:
In the left leg of the upper $g$ the subterm $g\0\Om$ (of $C$) is replaced by the subterm
demanding the first argument of $f$.
In the right leg the subterm $g\Om\0$ (of $C$) is replaced by the subterm demanding the
second argument of $f$.
This ensures that not both legs (of the upper $g$) can be evaluated.
There is again no term with $g$ that could be lifted to the top level and that would
eliminate the token $\bo\mto\{\bo\bo\mto\bz\}\mto\0$.
Therefore there is no $\lsy$-step leading from $A$ to $B$. 
But the subterms with $f$ can be lifted to the top replacing the upper $g$ of $B$
(as ``separator'' of $g\1\2$ and $g\1\1$), so we get $A$ with that token eliminated.
Here the subterms $g\0\Om$ and $g\Om\0$ of the former example $C$ appear again;
they must appear to ensure that $A$ gets the first eight tokens of $B$
and ensure that not both legs of the upper $f$ can be evaluated.

\begin{prop}
Let $A,B$ be the game terms of grade $2$ above.
There are no game terms $A',B'$ of grade $2$ with $A'\lsy B'$ and $A'\eq A$, $B'\eq B$.
Then by the game term theorem there are no PCF-terms $A',B'$ with this property.
This refutes the chain conjecture.
\end{prop}
\proof
As game terms of grade $2$, $A'$ and $B'$ should be of the form $\la fg.S$,
where $S\typ \io$ is a game term possibly with the only free variables $f,g$.
We abbreviate $S[f:=M,g:=N]$ as $S[M,N]$.

\no
We use the terms of the proof of proposition \ref{p:ABC} as arguments for $g$:
\begin{align*}
Q &= \la xy.\caseum x\0(\caset y \bo\bo\1) & \tsem{Q} &= \{\0\bo\mto\0,\tsp \1\2\mto\1\} \\
R &= \la xy.\caseum y\0(\caseum x \bo\1) & \tsem{R} &= \{\bo\0\mto\0,\tsp \1\1\mto\1\} \\
P &= \la xy. \0   &  \tsem{P} &= \{\bo\bo\mto\0\}\\
\intertext{We use the following terms as arguments for $f$:}
Q' &= \la xy.\caseum x (\caseum y \0\bo) \1  & \tsem{Q'} &= \{\0\0\mto\0, \tsp \1\bo\mto\1\} \\
R' &= \la xy.\caseum y (\caseum x \0\bo) \1  & \tsem{R'} &= \{\0\0\mto\0, \tsp \bo\1\mto\1\} 
\end{align*}
The pairs $(Q',Q)$ and $(R',R)$ are compatible in the sense that their replacement into
the same integer term leads to compatible results.

We will prove that for any terms $S,S'$ of the form above:
\[ \text{If } S'\lsy S \text{ and } S[\Om,P]\ret\0,\quad S'[Q',Q]\ret\0,\quad S'[R',R]\ret\0, \text{ then } S'[\Om,P]\ret\0.\]
The proposition follows immediately from this claim.

The proof is by induction on the term $S$: The cases $S=\Om,\0,\1,\2$ are clear.\\
Let $S=\caset(gS_1 S_2)S_3 S_4 S_5$ and $S'\lsy S$ with
$S'=\caset(gS'_1 S'_2)S'_3 S'_4 S'_5$.
(The case $S'=\Om$ is clear.)\\
Assume the three conditions of the claim.\\
Let $(gS'_1 S'_2)[Q',Q] \ret q$ and
$(gS'_1 S'_2)[R',R] \ret r$,
both terms must converge to integer constants.\\
From the compatibility of $(Q',Q)$ and $(R',R)$ follows the compatibility of $q$ and $r$,
so either $q=r=\0$ or $q=r=\1$.\\
As $(Q',Q)$ and $(R',R)$ are compatible,
it cannot be $S'_2[Q',Q]\ret \2$ and $S'_2[R',R]\ret \1$.\\ 
Therefore $q=r=\0$.\\
Then we get $S_3[\Om,P]\ret\0$, $S'_3[Q',Q]\ret\0$, $S'_3[R',R]\ret\0$.\\
By the induction hypothesis for $S_3$ we conclude $S'_3[\Om,P]\ret\0$, hence $S'[\Om,P]\ret\0$.
This fulfills the claim.

\no
Now let $S=\caset(fS_1S_2)S_3S_4S_5$.\\
Then $S[\Om,P]\eq \Om$, so the claim is fulfilled.
\qed

The refutation of the chain conjecture shows that already for second-order types
the correspondence of stable and syntactic order is destroyed;
there seems to be no simple syntactic characterization of the stable order.
But certainly the two orders are related, but in which sense?
A weaker conjecture that is now open is the following:

\begin{conj}[Maximality Conjecture]\label{c:maximality}
Every PCF-term without $\Y$ that is syntactically maximal (\ie contains no $\Om$)
is also stably maximal.
\end{conj}

The existence of chains of any length suggests a kind of ``metric'' on
finite elements $a\lst b$:
If there is a chain between $a$ and $b$ of least length $n$,
then the distance of $a$ and $b$ is $n$.
If there is no chain, then the distance is $\infty$.
But it might be doubted if this is meaningful, 
or if a transition $A\lst B$ like the example above (without chain) should also be counted as some
kind of elementary step of finite distance.

The example $A\lst B$ above shows us that
the syntactic order $\lsy$ is not enough to give a syntactic description of the stable order;
there are more ``syntactic'' relations needed.
We can imagine that $A$ is produced from $B$ by ``forcing'' the upper $g$ in $B$ to be strict in
one of its two arguments,
so that the token $\bo\mto\{\bo\bo\mto\0\}\mto\0$ is eliminated.


We tentatively propose an improved chain conjecture with such a new syntactic relation of ``strictification''.
For this we have to extend PCF with a new operator.
The theory of this extension has still to be properly developed;
so all propositions in the rest of this section have the status of conjectures.

In \cite{Paolini} Luca Paolini extends PCF with two new operators,
one of them called $\strict$ of type $(\io \fun \io)\fun\io$.
Suppose the operational semantics is given by an evaluation procedure $\eval$.
Then $\strict$ obeys the rules:
\begin{align*}
\text{If } \eval(M\0)\conver \text{ and } \eval(M\bo)\diver \text{ then } \eval(\strict M) &= \0 \\
\text{If } \eval(M\0)\conver \text{ and } \eval(M\bo)\conver \text{ then } \eval(\strict M) &= \1 
\end{align*}
Here $X\conver$ means that $X$ evaluates to some integer constant, $X\diver$ is the negation.
Paolini also gives an effective evaluation for $\strict$.

We use instead a new constant  $\str \typ (\io\fun\io)\fun\io$ that is the ``strict half''
of $\strict$, \ie we have the only rule:
\[\text{If } \eval(M\0)=\0 \text{ and } \eval(M\bo)\diver \text{ then } \eval(\str M) = \0 \]
$\str$ can be expressed by a term with $\strict$,
but $\strict$ cannot be expressed by $\str$.
Note that our $\str$ is finite.
An effective evaluation could also be given for $\str$.
($\str M$ tests if $M\0$ evaluates to $\0$ and in this process checks if $M$ demands its argument $\0$.)

On the extended language (PCF+$\str$) the operational equivalence $\eq$ is defined
in the usual way by observation through program contexts.
It is extensional, \ie $M\eq N$ iff for all $M'\eq N'$ it is $MM'\eq NN'$.
There is a fully abstract semantics $\sem{\phantom{M} }$ given by equivalence classes of terms;
these equivalence classes are construed as functions.
These functions are stable; we can define a trace semantics $\tsem{\phantom{M} }$ in the usual way,
with the stable order $\lst$ as the inclusion relation on traces.
All denotations are monotonic \wrt the stable order $\lst$.

$\str$ has the trace semantics
\[ \tsem{\str} = \{\{\0\mto \0\}\mto \0\} \]
Note that the token 
$\{\0\mto \0\}\mto \0$
expresses the fact that the argument function $\{\0\mto\0\}$ is strict,
its argument $\0$ is needed.
Note that $\str$ is not monotonic \wrt the extensional order of PCF;
it is $\sem{\str} \{\0\mto\0\} = \0$,
but $\sem{\str} \{\bo\mto\0\} = \bo$.
It is 
\[ \tsem{\str} \sub \tsem{\la f. \pif f\0 \pthen \0 \pelse \Om} = \{\{\0\mto\0\}\mto\0,\; \{\bo\mto\0\}\mto\0\} \]

All semantic elements preserve compatibility in the following sense.
Let us define the relation $\coh$ of \defin{hereditary compatibility}
on denotations:
for integers it is $m\coh n$ if $m=\bo$ or $n=\bo$ or $m=n$.
For functions it is $f\coh g$ if for all $x \coh y$: $fx \coh gy$.
All our functions $f$ of (PCF+$\str$) have the property that
$f \coh f$.
Paolini's operator $\strict$ does not have it.

With $\str$ we can define functions $\strictify_n\typ \si_n\fun\si_n$,
where $\si_n=(\io\fun \ldots \fun \io\fun\io)$ with $n\geq 1$ arguments.
E.g. $\strictify_2\typ (\io\fun\io\fun\io) \fun (\io\fun\io\fun\io)$,
\[ \strictify_2 = \la gxy. \pif(\str[\la z. \pif g(\pif z \pthen x \pelse \Om)(\pif z \pthen y \pelse \Om)
   \pthen \0 \pelse \0])\pthen gxy \pelse \Om \]
$\strictify_2 gxy$ tests if $gxy$ converges and $g\bo\bo$ diverges, and outputs $gxy$ in this case.
So $\strictify_2 gxy$ ``forces'' $g$ to be strict in one of its two arguments.
If it is not, then the output is $\bo$.

Let us replace in the example term $B$ above the upper occurrence of $g$ by $(\strictify_2 g)$
to get a new term $B'$.
Then $\sem{B'} = \sem{A}$, in the semantics of (PCF+$\str$).
$A$ is a ``strictification'' of $B$.

If $M$ is a term of (PCF+$\str$), then $\unstr(M)$ is defined as the term $M$ with all
occurrences of $\str$ replaced by $\la f. \pif f\0 \pthen \0 \pelse \Om$.
So $\unstr(M)$ is a PCF-term and $M\lst \unstr(M)$, in the semantics of the extended language.

Now we can define our complementary ``syntactic'' relation.
\begin{defi}
Let $M,N$ be PCF-terms of the same type.
$M$ is a \defin{strictification} of $N$, written $M\lsys N$, if there is a (PCF+$\str$)-term $M'$
with $\sem{M}=\sem{M'}$ (in the semantics of (PCF+$\str$))
and $\sem{\unstr(M')}=\sem{N}$ (in the semantics of PCF).
\end{defi}
\no Note that for PCF-terms $M,N$:
$(M\lsys N \ifthen \sem{M} \lst \sem{N})$
and $(M\eq N \ifthen M\lsys N)$.

\begin{conj}[improved chain conjecture]\label{c:icc}
In PCF we have:
For all finite elements $a\lst b$ there is a sequence $(M_i)$ of terms with
$1\leq i \leq n$, $\sem{M_1}=a$, $\sem{M_n}=b$,
and for every $i<n$ it is $M_i\lsy M_{i+1}$
or $M_i\lsys M_{i+1}$.
\end{conj}

A proof of this conjecture would be non-trivial and should first be tried on second-order types.
(It might be that types higher than second-order need new higher-type strictness operators
that cannot be defined from $\str$.)
Perhaps the situation should first be clarified in the realm of (PCF+$\str$)
and a conjecture of this kind should be proved there.

Our (PCF+$\str$) is the ``weakest'' sequential extension of PCF with a control operator.
It is properly included in (PCF+$\strict$),
this in turn is included in (PCF+H), the sequentially realizable functionals of  John Longley \cite{Longley};
see section 9 in \cite{Paolini} for an overview of such extensions of PCF.
(PCF+H) is included in SPCF (mentioned in the introduction), which is no more extensional.
For all these extensions of PCF it would be interesting to give syntactic characterizations of the stable order.
First it should be clarified if all types are definable retracts of some lower order types,
as is the case for (PCF+H) and SPCF.
This could make the proofs easier,
as we will see for unary PCF in  the following section.

\section{Unary PCF}

Here we will prove Berry's conjectures for unary PCF,
with the aid of Jim Laird's results \Laird.
Unary PCF is the calculus of PCF without $\Y$ and with the only constant $\0$ and $\casez$-expressions.
Its semantics is given by the finite elements of $\finsz$ for all $\si$,
with the orders $\lex$ and $\lst$.

We first repeat the general closure properties of the $\finsi$, seen as embedded in the $\Ds$ of an f-model,
taken from lemma \ref{l:finites}, proposition \ref{p:propI} and theorem \ref{t:sub}.

\begin{prop}
The $\finsi$ are finite and downward closed \wrt $\lst$.\\
For $a,b\in \finsi$, $a\glbex b \in \finsi$ is the glb \wrt $\lex$ in $\Ds$ and $\finsi$.
For $a\cost b$ it is also the glb \wrt $\lst$.\\
For $a,b\in \finsi$ that are $\lex$-bounded in $\Ds$, $a\lubex b \in \finsi$ is the lub \wrt $\lex$ in $\Ds$ and $\finsi$.\\
For a finite set $X\sub \finsi$ that has a stable upper bound,
all minimal stable upper bounds of $X$ are in $\finsi$.
The extensional lub $\Lubex X$ is one of those.
If $X$ has a stable lub, then it is $\Lubex X$.
\end{prop}

To apply Laird's results on definable retractions,
we augment unary PCF with product types $\si \pro \ta$.
The constructs of the whole language are:\\
$\0\typ \io$, $\Om^\si\typ \si$, $x^\si\typ \si$\\
If $M\typ \ta$, then $\la x^\si.M\typ \si\fun\ta$.\\
If $M\typ\si\fun\ta$ and $N\typ\si$, then $MN\typ \ta$.\\
If $M,N\typ\io$, then $\casez MN\typ\io$.\\
If $M\typ\si$ and $N\typ \ta$, then $\lpro M,N\rpro \typ \si\pro\ta$.\\
If $M\typ \si\pro\ta$, then $\proi M \typ \si$ and $\prot M \typ \ta$.

The reduction rules are:\\
$(\la x.M)N\re M[x:=N]$\\
$\casez \0 M \re M$\\
$\proi\lpro M,N\rpro \re M$\\
$\prot\lpro M,N\rpro \re N$

This section needs the products only as auxiliary constructions for the first-order types that are the
targets of Laird's retractions.
In this section the underlying language is always the augmented unary PCF with products
if products are not explicitly excluded.

Laird defines in \Laird\  a categorical notion of \defin{standard model} of unary PCF together with
order-extensionality and partial extensional order at each type.
He defines \defin{parallel composition} as the function $f$ with
$f\lpro \bo,\bo\rpro = \bo$,
$f\lpro \bo,\0\rpro =  f\lpro \0,\bo\rpro = \0$,
$f\lpro \0,\0\rpro = \0$.
A model is \defin{universal} at type $\ta$ if every element of $\ta$ is the denotation of a term.

\begin{defi}[Laird, definition 3.4 in \Laird]
Given types $\si,\ta$, a \defin{definable retraction from} $\si$ \defin{to} $\ta$ (in a model $\mathcal{M}$)
(written $\Inj: \si\retract \ta : \Proj$ or just $\si\retract \ta$)
is a pair of (closed) terms $\Inj\typ \si\fun\ta$ and $\Proj\typ \ta\fun\si$ such that
$\sem{\la x. \Proj(\Inj x)} = \id$ in $\mathcal{M}$.
\end{defi}

\begin{lem}[Laird, lemma 3.10 in \Laird]\label{l:Laird}
For any type $\ta$ there is a natural number $n$ such that there is a definable retraction from $\ta$
to some binary product form of $(\io\fun\io)^n$;
the same retraction for any standard order-extensional model without parallel composition.
\end{lem}

\begin{thm}[Laird, theorem 3.11 in \Laird]\label{t:Laird}
Any standard model of unary PCF which is order-extensional and excludes parallel composition is universal.
\end{thm}

We can build the stable biorder model of unary PCF as a collection of bicpos $(\Es,\lex,\lst)$
for every type $\si$:
We start with $E^\io = \{\bo,\0\}$ and $\bo\lex\0$, $\bo\lst\0$.\\
$E^{\si\pro\ta} = E^\si \pro E^\ta$ with the usual $\lex$ and $\lst$.\\
$\Est$ is the set of stable and monotone functions $f\typ (\Es,\lex,\lst)\fun (\Et,\lex,\lst)$.
(If $x\lex y$ then $fx\lex fy$.
If $x\lst y$ then $fx\lst fy$.
If $x\cost y$ then $f(x\glbex y) = fx \glbex fy$.
Continuity conditions are not necessary as the domains are finite.)
$\Est$ is ordered by the usual $\lex$ and $\lst$.

$(\Es,\lex,\lst)$ is not only a bicpo,
but a distributive bicpo where the stable lub of two $\lst$-compatible functions is
defined pointwise, by proposition 4.7.10 in Berry's thesis \Berry.
(If $f\cost f'$, then $(f\lubst f')x = fx \lubst f'x$.)
Therefore the stable lub of two elements is also defined by union on traces.

The stable biorder model fulfills the conditions of theorem \ref{t:Laird}, therefore it is universal
(and fully abstract).
This means that $(\Es,\lex,\lst)$ is isomorphic to $(\finsz,\lex,\lst)$ for types $\si$ without products.
In the following the semantics of unary PCF-terms is always taken in the model $(E^\si,\lex,\lst)$.
All this proves Berry's first conjecture for unary PCF:

\begin{thm}[Laird \Laird]
For every type $\si$ without products, the structure $(\finsz,\lex,\lst)$ is a distributive bicpo
(hence also a bidomain as it is finite).\\
For $a,b\in \finsz$ with $a\cost b$, $a\lubst b$ is given by $\trace(a\lubst b) = \trace(a)\cup \trace(b)$
and this lub is taken pointwise for functions $a,b$.
\end{thm}

With the aid of Laird's definable retractions we can prove a strong form of Berry's second conjecture 
for unary PCF, based on the fact that it is valid for first-order types.
First we need two lemmas on the reduction.

\begin{lem}
The reduction $\re$ on unary PCF with products is confluent and strongly normalizing.
Therefore it has unique normal forms.
The normal form of a term of a type without products does not contain any product subterm.
\end{lem}
\proof
The confluence can be proved with the main theorem of \cite{Mueller:lconfluence},
see also \cite[theorem 10.4.15, page 576]{Bethke}:
The rules of $\re$ without the $\beta$-rule are confluent on the applicative terms (\ie the terms
without $\la$), as they are orthogonal;
they are left-linear and not variable-applying.
Therefore their combination with the $\beta$-rule is confluent.

For the proof of strong normalization there seems to be no theorem in the literature
that would provide an easy modular check for the simply typed $\la$-calculus with
algebraic rewrite rules of our form.

Therefore we take the proof of strong normalization of the simply typed $\la$-calculus 
with products in the textbook \cite[chapter 6]{Girard/Taylor}
for the only atomic type $\io$ and augment it by the constant $\0$ and $\casez$-expressions.
The proof stays literally the same.
The only thing we have to add is a proof that if $M,N$ are strongly normalizable,
then $\casez MN$ is so; in the proof that all terms are reducible.
\qed

\begin{lem}\label{l:nf}
Let $\om$ be the following map on unary PCF-terms (where $n,m\geq 0$):
{\allowdisplaybreaks
\begin{align*}
\om(\laxn \0) ={} & \laxn \0 \\ 
\om(\laxn yM_1\ldots M_m) ={} & \laxn y\, \om(M_1) \ldots \om(M_m) \text{, for $y$ variable}\\
\om(\laxn \casez MN) ={} & \laxn \casez \om(M)\,\om(N), \\*
 &\text{if }\om(M)=\casez\ldots\text{ or }\om(M)=y\ldots\text{ with $y$ variable} \\
\om(\laxn \lpro M,N \rpro) ={} & \laxn \lpro \om(M),\om(N)\rpro \\
\om(\laxn \proi M) ={} & \laxn \proi\, \om(M), \text{ if }\om(M)=y\ldots\text{ with $y$ variable} \\
\om(\laxn \prot M) ={} & \laxn \prot\, \om(M), \text{ if }\om(M)=y\ldots\text{ with $y$ variable} \\
\om(M) ={} & \Om \text{, in all other cases}
\end{align*}}
$\om(M)$ is a normal form prefix of $M$, it pushes $\Om$s upwards.\\
If $M$ is a normal form, then $\om(M) \eq M$.\\
If $M\ret N$, then $\om(M)\lsy \om(N)$.\\
If $M\lsy N$, then $\om(M)\lsy \om(N)$.\\
We define $\nf(M) = \om(\text{the normal form of }M)$.\\
For all $M\lsy N$ it is $\nf(M)\lsy \nf(N)$.
\end{lem}
\proof
The first four propositions are clear, we prove here the last one;
the proof is similar to the one of lemma \ref{l:approx2}.

Let $M',N'$ be the normal forms of $M,N$.\\
As the reduction rules for $\re$ do not involve $\Om$,
all the reductions $M\ret M'$ can also be done in $N$.
(If $A\lsy B$ and $A\re A'$, then there is $B'$ with $B\re B'$ and $A'\lsy B'$.)\\
So there is $N''$ with $N\ret N''$ and $M'\lsy N''$.\\
By confluence of $\re$ it is $N''\ret N'$.\\
Then we get $\nf(M)=\om(M')\lsy\om(N'')\lsy\om(N')=\nf(N)$.
\qed

\begin{thm}
For every type $\si$ without products, for every $a\in \finsz$ there is a game term $A\typ \si$ with
$a=\sem{A}$ such that for every $b\lst a$ there is $B\lsy A$ with $b=\sem{B}$.
\end{thm}
\proof
By Laird's lemma \ref{l:Laird} there is a number $n$ and a definable retraction
$\Inj \typ \si \retract \ta\typ \Proj$, with $\ta$ some binary product form of $(\io\fun\io)^n$.\\
Let $A'$ be a term for $a$, $\sem{A'}=a$.\\
Let $A'' =\nf(\Proj(\Inj A'))$.
$A''$ does not contain any subterm of product type.\\
By the game term theorem \ref{t:gameterm} we get the desired game term $A=\gtz(\FP^\si_0 A'')$ with $A\eq A''$,
so $\sem{A}=a$.

Let $C=\nf(\Inj A')$.
$C=\lpro C_1,\ldots,C_n \rpro$ in some binary pair form,
where $C_i\eq\la x.\bo$ or $\la x.\0$ or $\la x.x$.\\
Let $b\lst a$. Then $\sem{\Inj} b \lst \sem{\Inj} a = \sem{C}$.\\
For every $i$, if $x\lst \sem{C_i}$ then $x=\sem{C_i}$ or $x=\bo$.
Therefore there is $B'\lsy C$ with $\sem{B'} = \sem{\Inj} b$.\\
Let $B''=\nf(\Proj B')$.
It is $A''=\nf(\Proj C)$. Therefore $B''\lsy A''$.\\
By the game term theorem \ref{t:gameterm} there is a game term $B=\gtz(\FP^\si_0 B'')$ 
with $B\eq B''$ and $B\lsy A$.\\
We have $b=\sem{\Proj}(\sem{\Inj}b)=\sem{\Proj}\sem{B'}=\sem{B}$.
\qed

\textbf{Remark:}
Please note that Laird's retractions are incredibly intelligent,
because they must introduce in the term $A'' = \nf(\Proj(\Inj A'))$
some nestings of variables that were not present in $A'$, to fulfill the proposition of the theorem.

It is a nice exercise (of three pages) to compute an example:
Take $\si=(\io\fun\io\fun\io)\fun\io$ and $A'=\la g.g\0\0\typ \si$.
The trace of $A'$ is
\[ \tsem{A'}=\{\{\bo\bo\mto\0\}\mto\0,
\{\0\bo\mto\0\}\mto\0,
\{\bo\0\mto\0\}\mto\0,
\{\0\0\mto\0\}\mto\0\}. \]

Going through Laird's proof of lemma \ref{l:Laird}, we get complicated terms
$\Inj\typ\si\retract\ta\typ\Proj$ with
$\ta=(((\io\fun\io)\pro(\io\fun\io))\pro(\io\fun\io))\pro((\io\fun\io)\pro\io)$.\\
We compute the normal forms:
\begin{align*}
 \Inj A' \ret C &= \lpro\lpro\lpro\la x.x,\la x.x\rpro,\la x.x\rpro,\lpro\la x.x,\underline{\0}\rpro\rpro\\
 \Proj(\Inj A')\ret A'' &= \la g.\casez[g(g\0(g\underline{\0}\0))\0] [g\0(g\underline{\0}\0)]
\end{align*}
This term is much more expanded than needed.

If we replace
the underlined $\underline{\0}$ in $C$ by $\Om$, we get a term $A''$ with both underlined $\underline{\0}$
replaced by $\Om$.
The trace of this new term $A''$ is
$\{\{\bo\bo\mto\0\}\mto\0,
\{\0\bo\mto\0\}\mto\0,
\{\bo\0\mto\0\}\mto\0\}$.
Note that there was no syntactically lesser term than $A'$ with this trace.

\medskip
\textbf{Remark:}
Another recommended exercise for the reader is to encode our first counter-example
(to Berry's second conjecture) of subsection \ref{s:firstexample} in unary PCF.
The booleans are encoded by the type $\beta=\io\fun\io\fun\io$ as usual.
The value $\0$ is represented by $\la xy.x$, $\1$ is represented by $\la xy.y$.
There are three more inhabitants of $\beta$: $\Om$, $\la xy.\casez xy$ and $\la xy.\0$.
The example is now of type $(\beta\fun\beta\fun\beta)\fun\beta$.
The term $D$ can be given an expanded form such that $A\lsy B=C\lsy D$.
In $D$ the top boolean $\la xy.\0$ is used (in one position) as the lub of $\la xy.x$ and $\la xy.y$.

\section{Outlook}

We have seen one trick to produce several examples which show that the stable order in PCF 
is not so regular as Berry had expected.
These counter-examples have as necessary ingredients:
at least two incompatible values and at least a second-order type with at least arity two of some
functional parameter.
To be precise, we still have to show that Berry's conjectures are valid in all
second-order types with functional parameters of only arity one,
see conjectures \ref{c:arityfirst} and \ref{c:aritysecond}.

With the refutation of the chain conjecture in section \ref{s:chain}
we have shown that there is no simple characterization of the stable order in terms of 
the syntactic order.
In fact the counter-example shows that there is not only the syntactic order that causes the stable order,
but that there are other syntactic relations needed with this property.
Such another relation was identified as the relation of ``strictification'',
and an improved chain conjecture \ref{c:icc} was tentatively proposed.

There should be some kind of full syntactic account of the stable order,
at least for second-order types.
For any type there should be syntactic conditions that are necessary for the relation $A\lst B$ of terms.
These should at least prove the maximality conjecture \ref{c:maximality}:
Every PCF-term without $\Y$ that is syntactically maximal is also stably maximal.

It would also be interesting to find syntactic characterizations of the stable order in
extensions of PCF by sequential control operators,
\ie in (PCF+$\str$), (PCF+$\strict$), (PCF+H) and SPCF,
see the remarks at the end of section \ref{s:chain}.

In this paper we have treated the problem of the syntactic characterization of the stable order,
but Berry originally had in mind the semantic characterization of the syntactic order.
In the light of the results of this paper this seems to be a problem of similar difficulty.
One should first seek necessary conditions for the syntactic order that are stronger than
the stable order.


\section*{Acknowledgement}
I thank Reinhold Heckmann for carefully reading drafts of this paper and many discussions.
I thank Reinhard Wilhelm and the members of his chair for their support.
I thank the anonymous referees for their valuable suggestions.
\bibliography{bib}
\bibliographystyle{plain}
\vspace{-30 pt}
\end{document}